\documentclass[acmsmall,screen]{acmart}
\pdfoutput=1

\def\BibTeX{{\rm B\kern-.05em{\sc i\kern-.025em b}\kern-.08emT\kern-.1667em\lower.7ex\hbox{E}\kern-.125emX}}

\setcopyright{acmcopyright}
\acmJournal{TOMS}
\acmYear{20xx}\acmVolume{0}\acmNumber{0}\acmArticle{0}\acmMonth{3}
\acmDOI{xx.xxxx/xxxxxx}

\usepackage{amsmath}             % for equation typesetting
\usepackage{amssymb}             % for equation typesetting
\usepackage{hhline}
\usepackage[algo2e,ruled]{algorithm2e}
\usepackage{subcaption}

\usepackage[capitalize]{cleveref}
\crefname{figure}{Fig.}{Figs.}
\Crefname{figure}{Figure}{Figures}
\crefname{table}{Tab.}{Tabs.}
\Crefname{table}{Table}{Tables}
\crefname{equation}{Eq.}{Eqs.}
\Crefname{equation}{Equation}{Equations}
\crefname{section}{Sec.}{Secs.}
\Crefname{section}{Section}{Sections}
\crefname{algocf}{Alg.}{Algs.}
\Crefname{algocf}{Algorithm}{Algorithms}

\newcommand{\su}[1]{\mathrm{SU}({#1})}
\newcommand{\so}[1]{\mathrm{SO}({#1})}
\newcommand{\HQ}[0]{\mathbb{H}}

\newcommand{\MG}[3]{\mathbb{M}_{{#1},{#2}}(\mathbb{#3})}
\newcommand{\MGsq}[2]{\mathbb{M}_{{#1}}(\mathbb{#2})}
\newcommand{\VG}[2]{\mathbb{V}_{{#1}}(\mathbb{#2})}

\newcommand{\conj}[1]{\overline{#1}}
\newcommand{\conjt}[1]{{#1}^*}
\newcommand{\trans}[1]{{#1}^T}

\newcommand{\cmplxrep}[1]{{#1}_\mathbb{C}}

\newcommand{\vc}[1]{\vec{\lowercase{#1}}}
\newcommand{\subrow}[2]{#1_{({#2})}}
\newcommand{\subcol}[2]{#1^{({#2})}}
\newcommand{\subsubrow}[3]{#1_{({#2},{#3})}}
\newcommand{\subsubcol}[3]{#1^{({#2},{#3})}}
\newcommand{\subrowcol}[3]{#1_{({#2})}^{({#3})}}
\newcommand{\singrow}[2]{\vc{#1}_{({#2})}}
\newcommand{\singcol}[2]{\vc{#1}^{({#2})}}

\begin{document}

\title[High-Performance Quaternion Matrix Multiplication]{On the Efficacy and High-Performance Implementation of Quaternion Matrix Multiplication}

\author{David B. Williams--Young}
\authornote{Corresponding Author}
\email{dbwy@lbl.gov}
\affiliation{%
  \institution{Lawrence Berkeley National Laboratory}
  \city{Berkeley} 
  \state{California}
  \postcode{94720}
}

\author{Xiaosong Li}
\email{xsli@u.washington.edu}
\affiliation{%
  \institution{University of Washington}
  \city{Seattle} 
  \state{Washington}
  \postcode{98195}
}

\begin{abstract}
Quaternion symmetry is ubiquitous in the physical sciences. As such, much work has been
afforded over the years to the development of efficient schemes to exploit this symmetry
using real and complex linear algebra. Recent years have also seen many advances in the
formal theoretical development of explicitly quaternion linear algebra with promising
applications in image processing and machine learning. Despite these advances, there
do not currently exist optimized software implementations of quaternion linear algebra.
The leverage of optimized linear algebra software is crucial in the achievement of
high levels of performance on modern computing architectures, and thus provides a
central tool in the development of high-performance scientific software. In this work,
a case will be made for the efficacy of high-performance quaternion linear algebra software
for appropriate problems. In this pursuit, an optimized software implementation of quaternion
matrix multiplication will be presented and will be shown to outperform a vendor tuned
implementation for the analogous complex matrix operation. The results of this work pave the
path for further development of high-performance quaternion linear algebra software which
will improve the performance of the next generation of applicable scientific applications.
\end{abstract}




\maketitle

\section{Introduction}
\label{sec:intro}

In the ever evolving ecosystem of high--performance computing
(HPC), the full exploitation of contemporary computational
resources must constitute a central research effort in
computationally intensive fields such as scientific computing.
However, it is often the case that simply applying conventional
algorithms and data structures to existing problems will yield
sub-optimal time--to--solution and resource management on modern
HPC systems. By exploiting the symmetry of a particular problem and
explicitly considering the structure and resources of these
computational architectures, one can often develop more optimal
computational research pathways.  Such development has the
potential to enable routine inquiry and simulation of systems which
were inaccessible or impractical by existing computational methods.

In this work, we consider the computational benefits of exploiting
the scalar and linear algebras generated by the quaternion numbers.
The quaternions, also known as Hamilton's quaternions, are a
hyper-complex number system which extends the complex numbers and
are isomorphic with the special unitary group, $\su{2}$
\cite{Hamilton1866_Quaternions}. Perhaps the most notable feature
of the quaternion numbers is the loss of scalar commutivity under
multiplication. As such, the algebra which is generated by the
quaternions is peculiar in that it more closely resembles that of
matrix algebra than that of its real or complex counterpart. Since their
inception, the quaternions have seen extensive application both in
pure
\cite{Cayley1845,Frobenius1878_book,Hurwitz22_MA1,Hopf31_MA637,Chevalley96_book}
and applied mathematics
\cite{Grubin70JSR1261,Deavours73_AMM995,Sudbery79_proc,Arnol95_RMS1,Palacios06CMDA239}.
Historically, the quaternions have been applied most successfully
in the treatment of rigid body mechanics due to their relationship
with $\su{2}$ (and thus the group of spacial rotations, $\so{3}$)
\cite{Grubin70JSR1261,Palacios06CMDA239}.  This application has
been widespread in the field of computer graphics to accelerate
video animation via algorithms such as Slerp
\cite{Shoemake85_ACM245}. From a computational perspective,
quaternion arithmetic offers an attractive alternative to complex
arithmetic in that it admits a higher arithmetic intensity and
smaller memory footprint than that of the the complex matrix algebra it
represents.  A demonstration of this state of affairs will be given
in the body of this work.

In the context of linear algebra, quaternions and quaternion linear
algebra naturally manifest in many scientific applications such as
quantum chemistry and nuclear physics
\cite{Gropen97_MP937,Saue00_3996,Helgaker02_JCC814,Jensen05_JCP114106,Carravetta06_PRA022501,Liu09_IJQC2167,Armbruster17_JCP054101,Nakai17_IJQCe25356,Repisky18_JCP204104}.
Traditionally, quantum mechanics and quantum field theories have been
formulated in terms of the complex numbers. 
However, since the earliest days of the development of quantum
mechanics, it has been known that the spinor nature of the
fermionic wave function (i.e. electrons, neutrinos, etc) admits a quaternion representation due
to its relationship with $\su{2}$
\cite{Adler95_book,Horwitz84_432}. Typically, this representation
manifests as a result of time--reversal being a global
symmetry of the Hamiltonian \cite{Wilkinson84_JCP278,Paldus03_book,Gropen97_MP937}.  
As such, the quaternion algebra has been exploited in the
development of efficient real \cite{Wilkinson84_JCP278} and complex
\cite{Shiozaki17_5} eigensolvers for time--reversal symmetric
Hamiltonians.  Despite the success and power of these
methods, their exploitation of the quaternion algebra is
\emph{implicit} in that the final computer implementation of these
methods is done in either real or complex matrix arithmetic; thus they
cannot leverage the full computational potential of the quaternion
arithmetic.  In this work, a case will be made for \emph{explicit}
exploitation of quaternion arithmetic in high--performance
software.

In general, high--performance methods for scientific applications
rely on highly tuned numerical linear algebra software
libraries which implement the  BLAS 
\cite{Krogh79_TOMS308,Hanson88_TOMS18,Duff90_TOMS18,Whaley02_TOMS135} 
and LAPACK \cite{LAPACK}  standards for performance on
modern HPC systems. Historically, numerical linear algebra
has been the archetypal example and
motivation for the careful consideration of a computer's
architecture in the development of high--performance
software \cite{Zubair94_IBM563,Dongarra98_SC38,Dongarra01_PC3,Geijn01_ICCS51,Geijn08_TOMS3}. 
It was realized early on that straightforward
implementations of operations such as matrix multiplication will
yield sub-optimal performance results and that achieving peak
performance on modern architectures requires a drastic
departure from conventional implementations. We refer the 
reader to the work of \cite{Geijn08_TOMS3} for a
reasonably contemporary discussion on the optimization of BLAS
functions, specifically matrix-matrix multiplication, on modern architectures. 
BLAS and LAPACK optimization still constitutes a major research effort in the field of 
numerical linear algebra, and this has led to a number of different approaches
which are available in open source \cite{Dongarra98_SC38,Yi13_ICHP1,Yunquan12_IEEE684,BLIS1,BLIS2,BLIS4,BLIS5} and
vendor tuned  software such as the Intel Math Kernel Library (MKL) and the IBM Engineering Scientific Subroutine Library (ESSL).

We note that over the years, there have been many important theoretical
developments in the field of quaternion linear algebra
\cite{Zhang97_LAIA21,Rodman14_book}. Many of necessary algorithmic
building blocks for ubiquitous operations such as eigenvalue and
singular value decompositions, and matrix factorizations have been
developed
\cite{Mehrmann89_NM83,Zhang97_LAIA21,Baker99_LAIA303,Opfer03_BITNM991,Loring12_EM250,Chen18_JCAM26,Zhao19_JCAM59}.
Such explorations have been key in the development of recent
methods for efficient signal and image processing by exploiting
the quaternion algebra
\cite{Sangwine03_ICIP809,Mars04_SP1177,Sangwine14_book,Shu16_NC416}.
Despite these successes, the field of quaternion linear algebra is
still in its infancy in terms of software adoption by scientists
and engineers. This is primarily due to the fact that, relative to
its  complex counterpart, there do not currently exist highly
optimized software implementations of quaternion linear algebra.
In order for quaternion linear algebra to become a viable
alternative to  complex linear algebra, such software must be
developed.  In this work, it will be demonstrated that optimized
implementations of quaternion linear algebra operations hold the
potential for leveraging drastic performance improvements relative
to analogous complex operations in problems which they may be
applied.

A key building block of optimized linear algebra algorithms is
that of an optimized implementation of matrix multiplication,
which we will refer to as GEMM in this work. Having an optimized
GEMM implementation is a necessary (but not necessarily
sufficient) condition for optimization more involved linear
algebra operations such as eigenvalue decomposition and matrix
factorization. As such, this work will focus on the performance
and implementation of a quaternion GEMM to demonstrate the
efficacy of high-performance quaternion linear algebra.

The remainder of this work will be organized as follows.
\Cref{sec:quaternion} will review the necessary theory for the
development of quaternion linear algebra and how one might
leverage this algebra as an alternative to complex linear algebra
for a special class of complex matrices.  \Cref{sec:matmult} will
briefly review the nature and abstract structure of
high--performance GEMM implementations.
\Cref{sec:matmult_quaternion} details an implementation of a
high-performance GEMM operation using explicitly quaternion arithmetic on
a contemporary computing architecture.  Finally, performance
results for the implementation quaternion GEMM relative to a
vendor tuned complex implementation will be presented in
\cref{sec:results}, and \cref{sec:conclusions} will provide an
examination of the future research which will be enabled as a
result of our findings.

\subsection{Notation}
\label{sec:notation}

Throughout the remainder of this work, we will adopt the following notation
conventions:
\begin{enumerate}

  \item Scalars of a ring $\mathbb F$ will be denoted with lower
  case letters $a \in \mathbb F$.

  \item $\MG{M}{N}{F}$ will denote the ring of $M$-by-$N$ matrices
  over $\mathbb F$. Further, $\MGsq{N}{F} \equiv
  \MG{N}{N}{F}$ for brevity. Matrices will be denoted with
  capital letters, $A \in \MG{M}{N}{F}$.

  \item For $A\in\MG{M}{N}{F}$, $\conj{A}$ will denote the
  conjugate of $A$, $\trans{A}$ its transpose, and $\conjt{A} =
  \trans{\conj A}$ will denote its conjugate transpose. For $a\in
  \mathbb F$, we note that $\conj{a}\equiv\conjt{a}$.

  \item For $N=1$, we denote the set of vectors over a ring as
  $\VG{M}{F}$, and denote its elements as lower-case letters,
  $\vc{a} \in \VG{M}{F}$.

  \item $A([\mu_1, \mu_2], :) \in \MG{\mu_2 - \mu_1}{K}{F}$
  represents the sub-matrix consisting of the $\mu_1$-st to
  $\mu_2$-nd rows of $A$. If $\mu_2-\mu_1$ is to be understood from
  the context, we use the abbreviated notation $\subrow{A}{\mu_1}
  \equiv A([\mu_1, \mu_2], :)$.

  \item $A(:, [\nu_1 , \nu_2]) \in \MG{M}{\nu_2 - \nu_1}{F}$
  represents the sub-matrix consisting of the $\nu_1$-st to
  $\nu_2$-nd columns of $A$. If $\nu_2-\nu_1$ is to be understood
  from the context, we use the abbreviated notation
  $\subcol{A}{\nu_1} \equiv A(:, [\nu_1 , \nu_2])$.

  \item The limiting cases of single row or column vectors will be
  denoted  $\singrow{A}{\mu} \equiv A(\mu, :)$ and
  $\singcol{A}{\nu} \equiv A(:, \nu)$.

  \item If the indices $\mu_3$ and $\mu_4$ are to be understood
  from the context, we denote the sub-matrix of a sub-matrix as
  $\subsubcol{A}{\mu_1}{\mu_3} =
  \subcol{A}{\mu_1}([\mu_3,\mu_4],:)$, and so on for row
  sub-matrices.
\end{enumerate}

\section{The Quaternion Algebra}
\label{sec:quaternion}

\subsection{Scalar Operations}
\label{sec:quaternion_scalar}

Fundamental to the development of quaternion linear algebra is the
nature of the scalar algebra generated by the quaternions. In this
work, the set of quaternion numbers will be denoted $\HQ$ and
is defined as the set of all $q$ that  \cite{Hamilton1866_Quaternions}
\begin{equation}
  q = q^0 e_0 + q^1 e_1 + q^2 e_2 + q^3 e_3, \qquad 
    \quad q^0, q^1, q^2, q^3 \in \mathbb R,
  \label{eq:quaternion_def}
\end{equation}
where $q^0$ is referred to as the \emph{scalar} component of $q$ and
$\{q^1,q^2,q^3\}$ is referred to as its \emph{vector} component.
$\{ e_0, e_1, e_2, e_3 \}$ are the quaternion basis elements
and they generate the skew--symmetric algebra defined by,
\begin{subequations}
\label{eq:quaternion_alg}
\begin{align}
  &e_0 e_j = e_j e_0 = e_j, \qquad j \in \{0,1,2,3\}, \\
  &e_i e_j = -\delta_{ij} e_0 + \sum_{k=1}^{3} \varepsilon_{ij}^k e_k,
    \qquad i,j \in \{1,2,3\},
\end{align}
\end{subequations}
where $\delta$ is the Kronecker delta and $\varepsilon$ is the 
totally anti-symmetric Levi-Civita tensor. As such, the following
must hold true
\begin{subequations}
\begin{align}
  &e_i e_j = - e_j e_i, \qquad i,j \in \{1,2,3\}, \, i\neq j, \\
  &e_1e_2e_3 = -e_0.
\end{align}
\end{subequations}
Given \cref{eq:quaternion_alg}, the product of quaternion
scalars $p,q\in\HQ$ is given by the Hamilton product
\begin{equation}
pq = \left( p^0q^0 - \sum_{i=1}^3 p^i q^i \right) e_0 +
     \sum_{k=1}^3 
       \left( p^0 q^k + p^k q^0 + 
         \sum_{i,j = 1}^3 \varepsilon_{ij}^k p^i q^j
       \right) e_k,
\label{eq:quaternion_prod}
\end{equation}
and is thus non-commutative
\begin{equation}
[p,q] \equiv pq - qp =  \sum_{i,j,k = 1}^3 \varepsilon_{ij}^k 
  \left( p^i q^j - p^j q^i \right) e_k.
\end{equation}

There are a number remarkable results that arise from the algebra
defined by \cref{eq:quaternion_def,eq:quaternion_alg} which are
important to the construction of quaternion linear algebra.
The first is that $\HQ$ constitutes a normed division algebra,
and is in fact the largest normed division algebra for which 
multiplication is associative \cite{Frobenius1878_book}. As such,
we may define a quaternion norm and inverse for every nonzero
element of $\HQ$,
\begin{subequations}
\label{eq:quaternion_da}
\begin{align}
  &\| q \| = \sqrt{\sum_{i=0}^3 \left( q^i \right)^2}, \\
  & q^{-1} = \frac{\conj{q}}{\| q \|} \quad \forall q \neq 0,
\end{align}
\end{subequations}
where we have defined the quaternion conjugate
\begin{equation}
  \conj{q} \equiv \conjt{q} = q^0 e_0 - q^1 e_1 - q^2 e_2 - q^3 e_3.
  \label{eq:quaternion_conj}
\end{equation}
We note here that quaternions of unit norm ($\| q \|=1$) are
known in the literature as \emph{versors} \cite{Hamilton1866_Quaternions}.
In examining \cref{eq:quaternion_da,eq:quaternion_conj}, the
expressions for norm, inverse and conjugate closely resemble
those of the complex numbers, $\mathbb C$. In fact, $\mathbb C$
is a sub-algebra embedded in  $\HQ$, and this relationship is crucial for
the development of the relationship between complex and quaternion linear algebra.

To examine the relationship between $\mathbb C$ and $\HQ$, we
introduce a common, simplified notation
\begin{equation}
  q = \underline{q}^0 + \underline{q}^1 e_2, \label{eq:quaternion_complex_rep}
\end{equation}
where
\begin{subequations}
\begin{align}
  &\underline{q}^0 = q^0 e_0 + q^1 e_1, \\
  &\underline{q}^1 = q^2 e_0 + q^3 e_1.
\end{align}
\end{subequations}
Consider the subset $\underline{\mathbb C} \subset \HQ$ defined
by
\begin{equation}
\underline{\mathbb C} = \left\lbrace  
  q\in\HQ \quad\vert\quad q^2 = q^3 = 0 \right\rbrace.
  \label{eq:complex_subalg}
\end{equation}
Note that
$\underline{q}^0,\underline{q}^1 \in \underline{\mathbb C}$.
The algebra defined by $\underline{\mathbb C}$ is exactly that of 
$\mathbb C$ (this may be easily verified through expansion of
\cref{eq:quaternion_prod}). Thus, 
$\underline{\mathbb C} \cong \mathbb C$ via the map
\begin{equation}
  \underline{q} = q^0 e_0 + q^1 e_1 
  \quad \longleftrightarrow \quad 
  z = q^0 + q^1 i,
  \label{eq:h_c_iso}
\end{equation}
with $\underline{q} \in \underline{\mathbb C}$ and $z \in \mathbb
C$. It is important to note
here that \cref{eq:h_c_iso} does not imply $e_0 = 1$ nor
$e_1 = i$, simply that there exists a bijection between 
$\mathbb C$ and $\underline{\mathbb C}$. Keeping this in mind,
however, it will typically be the case that one can use them 
interchangeably without ambiguity. As such, whether scalars of
the form given in \cref{eq:complex_subalg} are treated as
complex or quaternion scalars should will be implied from their
context in the following discussion.

While one tends to describe quaternions in terms of scalars (and
rightly so), the algebra which they generate more closely that of
a matrix algebra, specifically that of a Lie group
\cite{Hall15_Lie}. In the development of quaternion linear
algebra, it is instructive to examine the isomorphism between the
versors and the special unitary group $\su{2}$ through the mapping
of basis elements
\begin{equation}
  e_0 \mapsto \sigma_0, \quad 
  e_1 \mapsto i\sigma_3, \quad
  e_2 \mapsto i\sigma_2, \quad
  e_3 \mapsto i\sigma_1,
\end{equation}
where the Pauli matrices are given as
\begin{equation}
  \sigma_0 = \begin{bmatrix} 1 & 0 \\ 0 & 1 \end{bmatrix}, \quad
  \sigma_1 = \begin{bmatrix} 0 & 1 \\ 1 & 0 \end{bmatrix}, \quad
  \sigma_2 = \begin{bmatrix} 0 & -i \\ i & 0 \end{bmatrix}, \quad
  \sigma_3 = \begin{bmatrix} 1 & 0 \\ 0 & -1 \end{bmatrix}.
  \label{eq:pauli}
\end{equation}
By expanding in terms of the Pauli basis, it may be demonstrated
that the algebra of $\HQ$ is isomorphic to the algebra generated
by $\langle \su{2} \rangle \subset \MGsq{2}{C}$ via the map
\begin{equation}
q \in \HQ \quad \mapsto \quad 
\cmplxrep{q} =
%\begin{bmatrix}
%q^0 + iq^1  & q^2 + iq^3 \\
%-q^2 + iq^3 & q^0 - iq^1
%\end{bmatrix} =
\begin{bmatrix}
\underline{q}^0  & \underline{q}^1 \\
-\conj{\underline{q}}^{1} & \conj{\underline{q}}^{0}
\end{bmatrix}\in \MGsq{2}{C}. 
\label{eq:h_matrep}
\end{equation}
Here we have denoted the complex matrix representation of a quaternion scalar
with a subscript $\mathbb C$.

\begin{table}
\caption{Real floating point operations (FLOPs) comparison for elementary arithmetic
operations using $\HQ$ and $\MGsq{2}{C}$ data structures. Note that FLOP counts
for $\MGsq{2}{C}$ consider a generic complex matrix and assume no additional 
structure.}
\label{tbl:hc_flops}
\begin{tabular}{|l|c|c|}
\hline
Operation                          & FLOPs in $\HQ$ & FLOPs in $\MGsq{2}{C}$ \\ \hhline{|=|=|=|}
Addition       (\cref{eq:hc_add})  & 4              & 8                                \\
Multiplication (\cref{eq:hc_mult}) & 16             & 32                               \\
%Inversion      (\cref{eq:hc_inv})  & ?              & ?                                \\
\hline
\end{tabular}
\end{table}

While the representation given in \cref{eq:h_matrep} may seem inconsequential, it demonstrates
the great potential for improving computational performance by exploiting quaternion
arithmetic. As their algebras are isomorphic, it is known that
\begin{subequations}
\begin{alignat}{3}
&p + q \quad &&\longleftrightarrow \quad &&\cmplxrep{p} + \cmplxrep{q}, \label{eq:hc_add}\\
&pq \quad &&\longleftrightarrow &&\cmplxrep{p} \cmplxrep{q}, \label{eq:hc_mult}
\end{alignat}
\label{eq:hc_arith}
\end{subequations}
Although the result of both the quaternion and complex arithmetic may be 
thought of as to represent the same mathematical object (up to an isomorphism),
the computational work required for these operations is different for the
two arithmetics, respectively. \Cref{tbl:hc_flops} summarizes the number of 
\emph{real} floating point operations (FLOPs) required for the operations given in
\cref{eq:hc_arith}.
In this work we adopt the convention that the operation
\begin{equation}
a = a + bc, \qquad a,b,c\in\mathbb R,
\end{equation}
constitutes a single FLOP. From \cref{tbl:hc_flops}, we can see that there
is a 2x reduction in FLOPs for $\HQ$--arithmetic over generic 
$\MGsq{2}{C}$--arithmetic for the same mathematical operation.
We may further note that there is also a 2x reduction in memory operations
(MOPs) and computational storage requirements between $\HQ$ and 
$\MGsq{2}{C}$ data structures. This fact will prove important 
in the following developments of high--performance quaternion linear 
algebra.

As $\HQ$ forms a normed, associative division algebra, it is natural to consider
quaternion linear algebra, i.e. the algebra generated by matrices and vectors
of \emph{quaternion} elements, as an extension of the discussion presented in this subsection.
In the following subsection, we briefly review the relevant theory to motivate
the usage of explicitly quaternion linear algebra for software implementation.

\subsection{Matrices of Quaternions and Quaternion Linear Algebra}
\label{sec:quaternion_la}

Consider the space of $M\times N$ matrices with quaternion
elements denoted $\MG{M}{N}{H}$ and given by the generic element
$Q \in \MG{M}{N}{H}$ \cite{Zhang97_LAIA21},
\begin{equation}
Q = Q^0 e_0 + Q^1 e_1 + Q^2 e_2 + Q^3 e_3, \quad Q^0,Q^1,Q^2,Q^3 \in \MG{M}{N}{R}.
\label{eq:quaternion_mat_def}
\end{equation} 
Note the similarity of \cref{eq:quaternion_mat_def} with \cref{eq:quaternion_def}. 
In analogy with $\MG{M}{N}{C}$, we may define conjugate and conjugate transpose operations
on $\MG{M}{N}{C}$ via
\begin{subequations}
\begin{alignat}{2}
  &(\conj{Q})_{\mu\nu}  &&= \conj{Q_{\mu\nu}}, \\
  &(\conjt{Q})_{\mu\nu} &&= \conj{Q_{\nu\mu}}.
\end{alignat}
\end{subequations}
In the same manner as $\MGsq{N}{C}$, we define quaternion hermiticity as $Q = \conjt{Q}$.
Further, we may define a scalar and matrix product 
operations for $P\in\MG{M}{K}{H}$, $Q\in\MG{K}{N}{H}$ and $q \in \HQ$ \cite{Zhang97_LAIA21}.
\begin{subequations}
\begin{alignat}{2}
  &(qQ)_{\mu\nu} &&= qQ_{\mu\nu}, \\
  &(qPQ)_{\mu\nu} &&= q\sum_{\kappa=1}^K P_{\mu\kappa} Q_{\kappa\nu}.
\end{alignat}
\end{subequations}
However, unlike real and complex matrices, the loss of scalar commutivity
in $\HQ$ dictates that we must also consider operations of the form 
\begin{subequations}
\begin{alignat}{2}
  &(Qq)_{\mu\nu} &&= Q_{\mu\nu}q,  \\
  &(PqQ)_{\mu\nu} &&= \sum_{\kappa=1}^K P_{\mu\kappa} qQ_{\kappa\nu}, \label{eq:no_scalar_comm_mat}
\end{alignat}
\end{subequations}
where, in general, $qQ \neq Qq$ and $qPQ\neq PqQ\neq PQq$. In
addition, generally $PQ \neq QP$, however this is in perfect
analogy with real and complex linear algebra.  As one may
intuitively guess, this loss of scalar commutivity greatly
complicates proofs and algorithm development in quaternion linear
algebra \cite{Zhang97_LAIA21,Rodman14_book}, often requiring
researchers to resort to rather complex and abstract mathematical
paradigms, such as algebraic topology
\cite{Zhang97_LAIA21,Baker99_LAIA303}, to obtain the desired
outcomes.  Despite these complications, it is possible to extend
operations which are important to scientific application, such as
matrix inversion \cite{Zhang97_LAIA21,Loring12_EM250} and
eigenvalue decomposition
\cite{Zhang97_LAIA21,Mehrmann89_NM83,Baker99_LAIA303,Chen18_JCAM26,Zhao19_JCAM59},
to $\MGsq{N}{H}$.  In particular, the set of all invertable
quaternion matrices forms a group under the matrix product
\cite{Zhang97_LAIA21}.

Just as $\HQ$ admits a close relationship with $\mathbb{C}$ and $\MGsq{2}{C}$, 
analogous relationships may be developed
between $\MG{M}{N}{H}$, $\MG{M}{N}{C}$ and $\MG{2M}{2N}{C}$. Consider the
subset $\MG{M}{N}{\underline{C}} \subset \MG{M}{N}{H}$,
\begin{equation}
\MG{M}{N}{\underline{C}} = \left\lbrace  
  Q\in\MG{M}{N}{H} \quad\vert\quad Q_{\mu\nu}^2 = Q_{\mu\nu}^3 = 0 \right\rbrace.
  \label{eq:complex_subalg_mat}
\end{equation}
We may define an analogous expression to \cref{eq:quaternion_complex_rep} for
$\MG{M}{N}{H}$ via
\begin{subequations}
\begin{alignat}{2}
&Q               &&= \underline{Q}^0 + \underline{Q}^1 e_2,\\
&\underline{Q}^0 &&= Q^0 e_0 + Q^1 e_1,\\
&\underline{Q}^1 &&= Q^2 e_0 + Q^3 e_1,
\end{alignat}
\end{subequations}
with $\underline{Q}^0,\underline{Q}^1\in\MG{M}{N}{\underline{C}}$.
In the same manner as $\underline{\mathbb C}\cong \mathbb{C}$ (\cref{eq:h_c_iso}), 
$\MG{M}{N}{\underline{C}} \cong \MG{M}{N}{C}$ via the map
\begin{equation}
  \underline{Q} = Q^0 e_0 + Q^1 e_1 
  \quad \longleftrightarrow \quad 
  Z = Q^0 + Q^1 i.
  \label{eq:h_c_mat_iso}
\end{equation}
To construct its relationship to $\MG{2M}{2N}{C}$, we examine the 
$\MGsq{2}{C}$ representation of a quaternion matrix element,
\begin{equation}
  \cmplxrep{\left(Q_{\mu\nu}\right)} = 
    Q_{\mu\nu}^0 \sigma_0 + 
   iQ_{\mu\nu}^1 \sigma_3 + 
   iQ_{\mu\nu}^2 \sigma_2 + 
   iQ_{\mu\nu}^3 \sigma_1. 
  \label{eq:qmatel_complex}
\end{equation}
Thus \cref{eq:qmatel_complex} may be written in terms of a Kronecker product:
\begin{align}
  \cmplxrep{Q} &= 
    Q^0 \otimes \sigma_0 + 
   Q^1 \otimes i\sigma_3 + 
   Q^2 \otimes i\sigma_2 + 
   Q^3 \otimes i\sigma_1 \nonumber \\
  &=
  \begin{bmatrix}
    \underline{Q}^0 && \underline{Q}^1 \\
    -\conj{\underline{Q}}^{1} && \conj{\underline{Q}}^{0}
  \end{bmatrix} \in \MG{2M}{2N}{C},
  \label{eq:qmat_complex}
\end{align}
where we have denoted the complex matrix representation of the quaternion matrix with a subscript
$\mathbb{C}$ in analogy with \cref{eq:h_matrep}.

\begin{table}
\caption{Real floating point operations (FLOPs) comparison for common linear algebra
operations using $\MGsq{N}{H}$ and $\MGsq{2N}{C}$ data structures. As in \cref{tbl:hc_flops}, 
FLOP counts for $\MGsq{2N}{C}$ consider a generic complex matrix and assume no additional 
structure.}
\label{tbl:hc_mat_flops}
\begin{tabular}{|l|c|c|}
\hline
Operation                          & FLOPs in $\MGsq{N}{H}$ & FLOPs in $\MGsq{2N}{C}$ \\ \hhline{|=|=|=|}
Addition       (\cref{eq:hc_mat_add})  & $4N^2$                 & $8N^2$              \\
Multiplication (\cref{eq:hc_mat_mult}) & $16N^3$                & $32N^3$             \\
%Inversion      (\cref{eq:hc_mat_inv})  & ?              & ?                                \\
\hline
\end{tabular}
\end{table}

In analogy to \cref{eq:hc_arith}, the isomorphism between $\MG{M}{N}{H}$ and $\MG{2M}{2N}{C}$ 
admits the following relationships
\begin{subequations}
\begin{alignat}{3}
&P + Q \quad &&\longleftrightarrow \quad &&P_\mathbb{C} + Q_\mathbb{C}, \label{eq:hc_mat_add}\\
&PQ \quad &&\longleftrightarrow &&P_\mathbb{C} Q_\mathbb{C}. \label{eq:hc_mat_mult}
\end{alignat}
\label{eq:hc_mat_arith}
\end{subequations}
As in \cref{eq:hc_arith}, the amount of computational
work required to  perform the operations in \cref{eq:hc_mat_arith} in quaternion
and complex arithmetic are different. As an extension of \cref{tbl:hc_flops},
\cref{tbl:hc_mat_flops} summarizes differences in the the number of FLOPs required
for the same algebraic operation in the two arithmetics, respectively. For
simplicity and brevity, the summary in \cref{tbl:hc_mat_flops} only accounts 
for $\MGsq{N}{H}$ and $\MGsq{2N}{C}$, though completely analogous results hold
for the general rectangular case. Just as in \cref{tbl:hc_flops}, there is a 2x reduction
in both FLOPs and MOPs in utilizing explicitly quaternion arithmetic and data
structures over the analogous complex operations. However, this comparison is in terms of a
\emph{ratio} of computational work requirements. In terms of raw differences between
the two arithmetics, the potential computational savings scale to some power
of the dimension the matrix in question. For example, the difference in the number
of FLOPs required for quaternion / complex matrix multiplication is $16N^3$. For small
$N$ this would not make a drastic difference, but for large $N$, this difference becomes
significant. Due to the fundamental and central importance of matrix multiplication in
numerical linear algebra, similar comparisons could be made for any matrix operation,
such as eigenvalue decomposition or matrix factorization, between complex and quaternion
arithmetic.

\section{High--Performance Matrix--Multiplication}
\label{sec:matmult}

The cornerstone of high--performance numerical linear algebra software is
the optimized implementation of general matrix--matrix
multiplication (GEMM). Without an optimized GEMM implementation, operations such
as eigenvalue decomposition and matrix inversion
become impractical for large matrices. Thus,
the first step in the development of high-performance quaternion
linear algebra is the development of an optimized quaternion
GEMM. 

Over the past several decades, an enormous amount of research effort
in the fields of numerical linear algebra and HPC has been
directed towards the development of optimized GEMM operations on
various computing architectures. As a result, many different
strategies have been developed for high--performance GEMM
implementations 
\cite{Dongarra98_SC38,Geijn01_ICCS51,Geijn08_TOMS3,BLIS1,Yi13_ICHP1,Yunquan12_IEEE684}. 
Despite their differences, the common motif among these methods is
the rejection of a ``one--size--fits--all" development strategy
for all computing platforms, i.e. one must explicitly consider and
optimize for the underlying features of the computer
architecture in question to reach optimal performance. In
modern HPC, there are effectively three fundamental aspects of computing
architectures which must be considered in the development of
optimized GEMM operations \cite{Geijn08_TOMS3}:
\begin{enumerate}
  \item Efficient and effective utilization of various levels
  of the computational data and instruction caches.
  \item Utilization of microarchitechture specific features such as
  single instruction multiple data (SIMD) and fused multiply--add
  (FMA) operations.
  \item Achieving efficient parallelism on modern multi--core
  and many--core computing architectures.
\end{enumerate}
To demonstrate the efficacy of quaternion GEMM, we will only
consider the former two of these features; leaving the treatment of
parallelism for future work.
Due to its relative simplicity and portability to
general architectures, the development of high--performance
quaternion GEMM operations in this work will extend the strategy
adopted by the BLIS library for real and complex GEMM
operations \cite{BLIS1,BLIS2,BLIS4,BLIS5}. In the BLIS strategy, the
aspects of the GEMM operation which must be explicitly optimized for
a specific architecture are factored into a manageably small set of auxiliary procedures,
referred to as \emph{kernels},
while the general scaffold for the GEMM remains consistent between architectures.
Further, the structure and function of the kernels yielded by this strategy are 
designed in such a way that they may be used in the implementation of other  
BLAS-3 functionality such as rank-$k$ updates (XSYRK), triangular matrix multiplication 
(XTRMM), etc. 
In this section, we examine the salient
aspects of this strategy which are agnostic to the data
representation and arithmetic operations relating to the matrices being
multiplied.

\subsection{The General Algorithm}
\label{sec:matmult_general}

Consider the GEMM of two matrices $A\in\MG{M}{K}{F}$,
$B\in\MG{K}{N}{F}$ over a general ring $\mathbb F$,
\begin{equation}
  C = \alpha A B + \beta C, 
  \label{eq:general_gemm}
\end{equation}
where $\alpha,\beta\in\mathbb F$ and $C \in \MG{M}{N}{F}$.
Computationally, $A,B$ and $C$ are stored as linear, contiguous
data structures of lengths $MK$, $KN$ and $MN$, respectively. 
%In
%this work, we will consider both row-- and column--major
%storage of matrices.
%In column--major storage: $A_{\mu\kappa}$
%and $A_{(\mu+1)\kappa}$, and $A_{M\kappa}$ and $A_{1(\kappa+1)}$
%are stored contiguously in memory, 
%while in row--major storage: $A_{\mu\kappa}$
%and $A_{\mu(\kappa+1)}$, and $A_{\mu K}$ and $A_{(\mu+1)1}$
%are stored contiguously in memory. 
In
this work, we will consider column--major
storage of matrices, i.e.
$A_{\mu\kappa}$
and $A_{(\mu+1)\kappa}$, and $A_{M\kappa}$ and $A_{1(\kappa+1)}$
are stored contiguously in memory. 

\begin{figure}[t]
  \centering
  \includegraphics[width=0.3\textwidth]{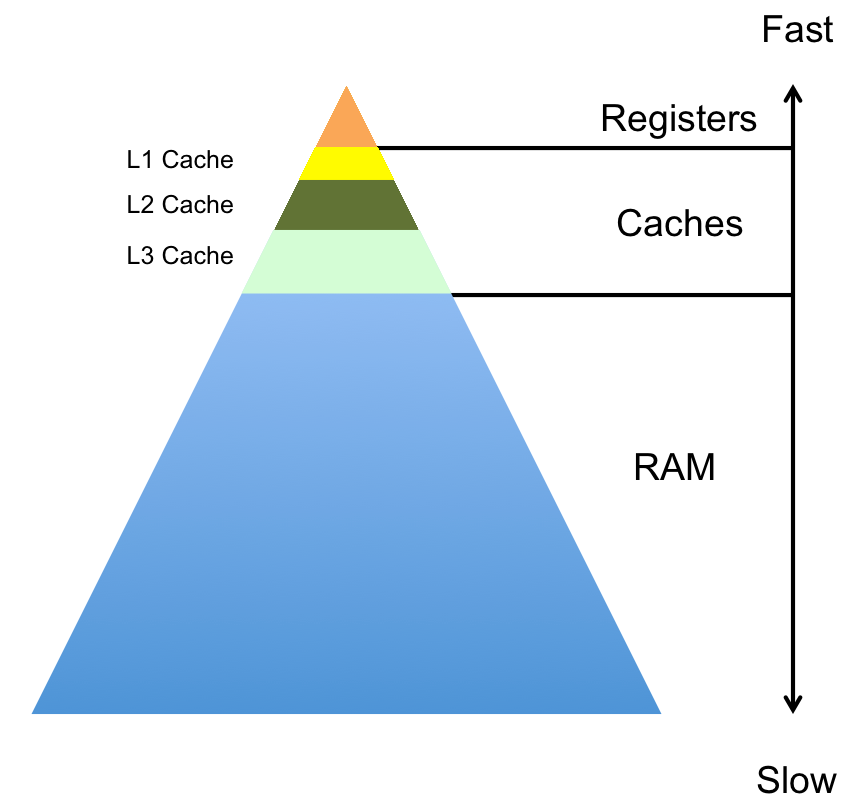}
  \caption{A simplified model of the memory hierarchy on
  modern computing architectures.}
  \label{fig:caches}
\end{figure}

\begin{algorithm2e}[b]
  \caption{Reference GEMM Algorithm}
  \label{alg:reference_gemm}
  \SetKwInOut{kinput}{Input}
  \SetKwInOut{koutput}{Output}
  \BlankLine
  \kinput{Matrices $A \in \MG{M}{K}{F}, B \in \MG{K}{N}{F}, 
          C \in \MG{M}{N}{F}$, \\ 
          Scalars $\alpha,\beta \in \mathbb F$ }
  \BlankLine
  \koutput{$C = \alpha AB + \beta C $}
  \BlankLine
  \For{$\nu = 1:N$}{
    %\BlankLine
    \nl Load $\singcol{C}{\nu} = C(:, \nu)$ into cache\\
    \nl $\singcol{C}{\nu} = \beta \singcol{C}{\nu}$ \\
    %\BlankLine
    \For{$\kappa = 1:K$}{
      %\BlankLine
      \nl Load $\singcol{A}{\kappa} = A(:, \kappa)$ into cache \\
      \nl Load $B_{\kappa\nu}$ \\
      \nl $\singcol{C}{\nu} =  \singcol{C}{\nu} + \alpha \singcol{A}{\kappa} B_{\kappa\nu} $
      %\BlankLine
    }
    %\BlankLine
    \nl Store $\singcol{C}{\nu}$
  }
\end{algorithm2e}

For comparison in the following, \cref{alg:reference_gemm}
outlines the simplest implementation of the GEMM operation which
was suggested in the earliest developments of the BLAS standard
\cite{Duff90_TOMS18}.  This method will be referred to as the
\emph{reference} GEMM algorithm. To fully understand the drawbacks
of \cref{alg:reference_gemm} and to motivate the development of a
more optimal algorithm, we must examine that nature of the memory
hierarchy on modern computers. \Cref{fig:caches} illustrates a
simplified model of a representative memory hierarchy on a modern
computer \cite{Geijn08_TOMS3}.  At the top of the hierarchy, the
fastest and least abundant memory resource are the registers which
physically reside on the processor.  It is on the data which
resides in the registers that the processor may issue instructions
such as arithmetic operations, etc.  On architectures which
support SIMD instructions, i.e vector processors with instruction
sets such as SSE, AVX, AVX2 and AVX-512, each floating point
register can hold a small number of floating point numbers at a
time, typically between 2 and 16.  However, their abundance is
very limited: 16 registers on SSE, AVX and AVX2, and 32 on
AVX-512. Thus, to fully exploit the speed of the registers, care
must be taken to carefully populate the data which resides there
to minimize the data movement between the registers and other
levels of the hierarchy.

At the bottom of the hierarchy is the slowest and largest memory
resource: the random access memory (RAM). It is in the RAM that
the matrices which participate in the GEMM operation are typically
stored. The RAM is the memory resource that resides furthest from
the processor, which allows it to be orders of magnitude larger
than any of the other memory resources (on the order of 1GB-1TB).
However, the penalty for its size and distance from the processor
is a high access latency.  The speed at which data can be moved to
and from RAM varies drastically between different architectures
and manufacturers,  and also depends on factors such as how the
data being moved is laid out in memory (contiguous, strided, cache
aligned, etc).  Generally, reading to and from RAM amounts to
hundreds of clock cycles on modern processors.  Due to its very
high latency, data movement in and out of RAM must be kept to a
minimum to achieve optimal performance. As such data is rarely
read directly from RAM to the registers or visa versa. Instead, it
is typically the case that data is read to and from the RAM to a
low latency intermediary storage, known as the data cache, which
resides closer to the processor and is thus capable of moving data
to and from the registers much faster than would be possible from
the RAM.

Due to the slow rate at which RAM may be accessed, typical memory
access patterns dictate that the RAM should be read in large
chunks of contiguous data into the cache. Whenever a read
instruction is issued from the processor for a particular memory
address in RAM, it first checks if that data resides in cache. If
the data resides in cache, what is referred to as a \emph{cache
hit}, it may be read directly from cache and avoid the RAM
completely. However, if the data does not reside in cache when the
instruction is issued, the data must be moved from RAM to cache
and then read into the registers. This process is referred to as a
\emph{cache miss}. Due to the large latency differential between
RAM and cache, the penalty for a cache miss can often be quite
large.  Further, the restricted size of the cache only allows a
limited amount of data can be stored there at any point in time.
When the cache reaches its capacity, data which resides in the
cache must be replaced when new data is read in from the RAM. The
process by which this replacement happens is referred to as the
cache's replacement policy.  If data is to be often reused in an
algorithm, it is important to ensure that it resides in the cache
as often as possible to minimize the probability of a cache miss.
Thus, knowledge of the replacement policy is paramount in the
development of a strategy for cache population to maximally reuse
the data the resides there while not ejecting reusable data with
data which is to be used less often.

On contemporary architectures, the cache is divided into cascading
``levels": the L1, L2, L3 caches, etc. The capacity and access
latencies for the cache levels vary considerably between processor
generations and manufacturers; however, the general trend is to
lose an order of magnitude on access latency and gain an order of
magnitude in capacity between successive cache levels. For
example, the Intel(R) Xeon(R) CPU E5-2660 (Sandy Bridge) processor
yields cache capacities of 32 kB, 256 kB, 20 MB and access
latencies of 4, 12 and 29 clock cycles for the L1, L2 and L3
caches, respectively \cite{Fog12_Micro}.  It is typically the case
that the population of the different levels of cache cannot be
explicitly programmed; one typically relies on heuristics issued
by the CPU, such as data prefetching and cache replacement
policies, to perform this population. However, with knowledge of
the sizes of the cache levels and replacement policies, one may
develop algorithms which aim to populate these caches optimally
for data reuse.

From the perspective of effective utilization of
the memory hierarchy and the other aforementioned features of
computing architectures, there are a number of 
drawbacks in \cref{alg:reference_gemm}:
\begin{itemize}
  \item All of $A$ is loaded into cache for each column of $C$,
  \item For large $M,K$, loading $A$ potentially ejects $\singcol{C}{\nu}$
  from cache, triggering a cache miss on each update of $\singcol{C}{\nu}$,
  \item There is no useful caching of $B$,
  \item In a high-level programming language, this algorithm
  relies on an optimizing compiler to utilize SIMD, FMA, etc.
  \item Scalable parallelism is non--trivial.
\end{itemize}
\Cref{alg:reference_gemm} is referred to as a \emph{memory bound}
algorithm, i.e. its performance is completely determined by the
latency at which data may be moved to and from the RAM. As such,
even for relatively small GEMM operations, performance will be
sub-optimal \cite{Geijn08_TOMS3}. A demonstration of this state of
affairs in the context of quaternion GEMM will be given in
\cref{sec:results}.

\begin{algorithm2e}
  \caption{General Layered GEMM Algorithm}
  \label{alg:opt_gemm}
  \SetKwInOut{kinput}{Input}
  \SetKwInOut{koutput}{Output}
  \BlankLine
  \kinput{Matrices $A \in \MG{M}{K}{F}, B \in \MG{K}{N}{F}, 
          C \in \MG{M}{N}{F}$, \\ 
          Scalars $\alpha,\beta \in \mathbb F$, \\
          Caching parameters $N_c, M_c, K_c\in\mathbb Z^+$,\\
          Register block sizes $N_r, M_r \in \mathbb Z^+$ }
  \BlankLine
  \koutput{$C = \alpha AB + \beta C $}
  \BlankLine
  \BlankLine
  \BlankLine
  \nl Allocate $\tilde A_p\in\MG{K_cM_r}{M_c/M_r}{F}$, $\tilde B_p\in\MG{K_cN_r}{N_c/N_r}{F}$ \\
  \nl $C = \beta C$ \\
  \For{$\nu = 1:N:N_c$}{
    %\BlankLine
    \nl Identify $\subcol{C}{\nu} = C(:, [\nu,\nu+N_c])$ \\
    \nl Identify $\subcol{B}{\nu} = B(:, [\nu,\nu+N_c])$ \\
    %\BlankLine
    \For{$\kappa = 1:K:K_c$}{
      %\BlankLine
      \nl Identify $\subcol{A}{\kappa}         = A(:, [\kappa:\kappa+K_c])$ \\
      \nl Identify $\subrowcol{B}{\kappa}{\nu} = \subcol{B}{\nu}([\kappa,\kappa+K_c], :)$ \\
      \nl Pack $\tilde B_p \leftarrow $ \texttt{PACK2}$(\subrowcol{B}{\kappa}{\nu})$ (L3 cache) \label{alg_ln:b_pack}\\
      %\BlankLine
      \For{$\mu = 1:M:M_c$}{
        %\BlankLine
        \nl Identify $\subrowcol{C}{\mu}{\nu}    = \subcol{C}{\nu}([\mu,\mu+M_c], :)$ \\
        \nl Identify $\subrowcol{A}{\mu}{\kappa} = \subcol{A}{\kappa}([\mu,\mu+M_c], :)$ \\
        \nl Pack $\tilde A_p \leftarrow  \alpha * \texttt{PACK1}(\subrowcol{A}{\mu}{\kappa})$ (L2 cache) \label{alg_ln:a_pack}\\
        \nl $j_r \leftarrow 0$ \\
        \For{$\nu_r = 1:N_c:N_r$}{
          %\BlankLine
          \nl Identify $\subsubcol{\subrow{C}{\mu}}{\nu}{\nu_r} = \subrowcol{C}{\mu}{\nu}(:, [\nu_r,\nu_r+N_r])$ \\
          \nl Identify $\singcol{B_p}{j_r} = \tilde B_p(:, j_r)$ \\
          \nl $i_r \leftarrow 0$ \\
          %\nl Load $\tilde B_p^{(j_r)}$ into L1 cache \\
          %\BlankLine
          \For{$\mu_r = 1:M_c:M_r$}{
            %\BlankLine
            \nl Identify $\subsubrow{\subsubcol{C}{\nu}{\nu_r}}{\mu}{\mu_r} =\subsubcol{\subrow{C}{\mu}}{\nu}{\nu_r}([\mu_r,\mu_r+M_r], :) $ \\
            \nl Identify $\singcol{A_p}{i_r} = \tilde A_p(:, i_r)$ \\
            \nl $\subsubrow{\subsubcol{C}{\nu}{\nu_r}}{\mu}{\mu_r} \leftarrow $ 
             \texttt{KERN}($\subsubrow{\subsubcol{C}{\nu}{\nu_r}}{\mu}{\mu_r},
                           \singcol{A_p}{i_r}, 
                           \singcol{B_p}{j_r}$) \\ 
            \nl $i_r \leftarrow i_r + 1$\\
            %\BlankLine
          }
          \nl $j_r \leftarrow j_r + 1$\\
          %\BlankLine
        }
        %\BlankLine
      }
      %\BlankLine
    }
    %\BlankLine
  }
  %\BlankLine
  \nl Free $\tilde A_p, \tilde B_p$
\end{algorithm2e}

In order to overcome the memory bottle neck, one must develop an
algorithm which populates the levels of cache and registers with
sub-matrices of $A,B$ and $C$ according how their data may be
reused throughout the GEMM operation. For a detailed explanation
of the extent to which one may reuse different sub-matrices of
$A,B$ and $C$, we refer the reader to the work of
\cite{Geijn08_TOMS3}.  In general, the mechanism by which one
achieves optimal cache utilization is through a layered approach
to the GEMM operation
\cite{Dongarra98_SC38,Geijn01_ICCS51,Geijn08_TOMS3,BLIS1}.  An
optimized layered GEMM algorithm may be constructed through the
specification of three caching parameters: $M_c,N_c,K_c \in
\mathbb Z^+$, two register blocking parameters: $N_r, M_r \in
\mathbb Z^+$, two packing kernels: \texttt{PACK1}, \texttt{PACK2},
and a microkernel, \texttt{KERN}.  A representative example of
such an algorithm, specifically the algorithm which has been
proposed in the development of the BLIS framework
\cite{BLIS1,BLIS2,BLIS5}, is outlined in \cref{alg:opt_gemm}. For
simplicity in \cref{alg:opt_gemm}, we have assumed ${(N\mod N_c)} =
{(M\mod M_c)} = {(K\mod K_c)} = 0$ and ${(N_c \mod N_r)} = {(M_c
\mod M_r)} = 0$.  However, extension of \cref{alg:opt_gemm}
without these constraints is straightforward through zero padding
in the packing kernels \cite{BLIS1}.  We note for clarity that the
scaling by $\alpha$ in \cref{alg_ln:a_pack} of \cref{alg:opt_gemm}
may instead be performed in \cref{alg_ln:b_pack} for rings
$\mathbb F$ which admit scalar commutivity in the sense of
\cref{eq:no_scalar_comm_mat} (i.e. $\mathbb R$ and $\mathbb C$).
Each of these parameters and kernels must be carefully chosen and
optimized for each computer architecture of interest. In the
following subsection, we examine the nature of each of these
moieties and the factors one must consider in their selection. 

\subsection{Register Blocking and The Microkernel}

Consider the expression of a specific sub-matrix  $C_r =
C([i_1,i_2], [j_1,j_2])$ in terms of the corresponding
sub-matrices $A_r = A([i_1,i_2], :)$ and $B_r = B(:, [j_1,j_2])$,
\begin{equation}
C_r =  \sum_{\kappa=1}^K \singcol{A_r}{\kappa} \singrow{B_r}{\kappa}. \label{eq:micro_kernel}
\end{equation}
In other words, $C_r$ may be expressed as a sum of rank--1
updates over rows and columns of $B_r$ and $A_r$,
respectively. As this is the fundamental arithmetic operation of
the GEMM operation to be performed by the CPU, $\singcol{A_r}{\kappa}, 
\singrow{B_r}{\kappa}$ and $C_r$ must all
reside in the registers for the operation to take place.
$C_r$ is referred to as the \emph{register
block} of $C$, with dimensions $N_r = i_2 - i_1$ and $M_r = j_2 - j_1$. To
achieve optimal memory performance, $N_r$ and $M_r$ must be chosen
such that $\singcol{A_r}{\kappa}, \singrow{B_r}{\kappa}$
and $C_r$ may reside in the registers simultaneously in order
to avoid data movement between the registers and other levels of
the memory hierarchy \cite{Geijn08_TOMS3}.

\begin{algorithm2e}[t]
  \caption{Abstract Template for the Microkernel}
  \label{alg:abstract_kern}
  \SetKwInOut{kinput}{Input}
  \SetKwInOut{koutput}{Output}
  \BlankLine
  \kinput{ 
    Columns $\vc{a_p} \in \VG{K_cM_r}{F}$, 
    $\vc{b_p} \in \VG{K_cN_r}{F}$ of packed representations 
    $\tilde A_p\in\MG{K_cM_r}{M_c/M_r}{F}, \tilde B_p\in\MG{K_cN_r}{N_c/N_r}{F}$ of sub-matrices 
    $A_r \in \MG{M_r}{K_c}{F}, B_r \in \MG{K_c}{N_r}{F}$,
    respectively. \\
    Sub-matrix $C_r \in \MG{M_r}{N_r}{F}$.}
  \BlankLine
  \koutput{Partially updated $C_r$}
  \BlankLine
  \BlankLine

  \nl Load $C_r$ into registers.\\
  \For{$\kappa = 1:K_c$}{
    \nl Load $\singcol{A_r}{\kappa}$ and $\singrow{B_r}{\kappa}$  from $\vc{a_p}$ and $\vc{b_p}$ into registers.\\
    \nl $C_r \leftarrow C_r + \singcol{A_r}{\kappa} \singrow{B_r}{\kappa}$\\
  }
  \nl Store $C_r$.
\end{algorithm2e}
In \cref{alg:opt_gemm}, the full product, $C$, is constructed by
successively updating each of its (disjoint) $\MG{M_r}{N_r}{F}$
sub-matrices via partial summation (over $K_c$ elements) of
\cref{eq:micro_kernel} with the microkernel performing arithmetic
operations which amount to the sum over rank-1 updates. As the
arithmetic kernel of the GEMM operation, the microkernel is the
fundamental operation which is most sensitive to the underlying
computer architecture and is a key factor in the performance of
the GEMM implementation.  It is in the microkernel that one must
explicitly consider microarchitechture specific operations such as
SIMD and FMA. As such, optimized GEMM implementations typically do
not express the microkernel in a high-level language; it is
typically expressed directly in assembly language
\cite{Dongarra98_SC38,Geijn08_TOMS3,BLIS1} or with use of
low--level access paradigms such as vector intrinsics in C++. An
abstract template for a generic microkernel implementation is
given in \cref{alg:abstract_kern}.

There is a subtle, yet crucial aspect of the loop expressed in
\cref{alg:abstract_kern} in relationship to \cref{alg:opt_gemm}:
as all of the arithmetic intensity is folded into the rank-1
updates performed from within the microkernel inner-loop,
optimality of the GEMM operation is directly related to the amount
of time spent in this loop. In other words, the number of
operations performed inside of this loop, whether they be FLOPs or
MOPs, must be kept to a minimum to achieve optimal performance.
The number of FLOPs required to perform the rank-1 update is fixed
based on $M_r,N_r$ and $\mathbb F$, thus optimality is generally
achieved through minimizing the number of MOPs performed inside
this inner loop. To this end, the microkernel utilizes packed
representations, $\tilde A_p$ and $\tilde B_p$, of sub-matrices,
$A_r$ and $B_r$, produced by the packing kernels,
\texttt{PACK1} and \texttt{PACK2}, respectively.  The remainder of
this section is dedicated to the design and optimization of the
packing kernels and caching parameters to achieve optimal data
movement between levels of the memory hierarchy and to minimize
the number of MOPs required to be performed from within the
microkernel.

\subsection{Sub-matrix Packing for Optimal Data Layout and Cache Utilization}
\label{sec:packing}

Perhaps the most ingenious aspect of the layered GEMM algorithm
outlined in \cref{alg:opt_gemm} is the utilization of auxiliary
memory and packing kernels to amortize the cost of data
manipulation over the movement of data between the levels of the
memory hierarchy \cite{Geijn08_TOMS3}.  This packing strategy has
two primary objectives:
\begin{enumerate}
  \item To populate the various levels of the cache with
  sub-matrices of $A$ and $B$ according to the extent which they
  will be reused in the GEMM operation as to minimize probability
  of triggering cache misses,

  \item To ensure optimal, contiguous data layouts of the packed
  sub-matrices to minimize the number of operations (FLOPs and 
  MOPs) which must be performed from within the inner loop of the
  microkernel.
\end{enumerate}
In the following, we will examine both of these objectives in turn.

To optimize data movement for cache utilization, one must obtain
optimal choices for the caching parameters $M_c, N_c$ and $K_c$
for the architecture of interest. Typically, these parameters are
chosen such that \cite{Geijn08_TOMS3, BLIS1}: 
\begin{itemize}
  \item Contiguous storage of size $N_cK_c$ may reside in and be
  addressed from the L3 cache once the data is loaded from RAM 
  (e.g.  $\tilde B_p \leftarrow \subrowcol{B}{\kappa}{\nu}$) 
  until it is no longer needed.

  \item Contiguous storage of size $M_cK_c$ may reside in and be
  addressed from the L2 cache once the data is loaded from RAM 
  (e.g.  $\tilde A_p \leftarrow \subrowcol{A}{\mu}{\kappa}$) 
  until it is no longer needed.

  \item Contiguous storage of size $K_cN_r$ may be moved from the
  L3 to the L1 cache without triggering a cache miss or cache
  invalidation (e.g. $\singcol{B_p}{j_r} \leftarrow \tilde B_p$).
\end{itemize}
Clearly, the choice of these parameters are integrally tied to the
sizes of the L1, L2 and L3 caches and the size of the data
structure which represents $\mathbb F$. Several methods exist for
determining optimal choices for the caching parameters. There has
been work in the development of analytical models and formulas
which take into account the specifics of $\mathbb F$ and the
architecture in question and return optimal values for the caching
parameters \cite{BLIS4}. Other approaches utilize guided or
black--box optimization
\cite{Dongarra01_PC3,Yunquan12_IEEE684,Yi13_ICHP1}, to obtain
these parameters.  Once these parameters have been determined, the
task then becomes to develop efficient packing utilities which
optimize the data layout for use with the microkernel.

There are a number of desirable features one wishes to express in
the data layout of packed matrices, $\tilde A_p$ and $\tilde B_p$,
to optimize the data movement between the levels of cache and the
registers from within the microkernel:
\begin{itemize}
  \item The elements of $\singcol{A_r}{\kappa}$ and
  $\singrow{B_r}{\kappa}$ should be contiguous, respectively.
  As vectors, this amounts to ensuring $\vc{a}_{r,\mu}^{(\kappa)}$
  and $\vc{a}_{r,\mu+1}^{(\kappa)}$ are
  contiguous in memory, and similarly for $\singrow{B_r}
  {\kappa}$.

  \item The elements of $\tilde A_p$ and $\tilde B_p$ which
  contribute to adjacent register blocks of $C$ should be 
  contiguous in memory, i.e. $\singcol{A_p}{i_r}$ and 
  $\singcol{A_p}{i_r+1}$ should be contiguous in memory.

  \item For $\mathbb F$ which is represented by a compound
  datatype of primitive data, e.g. $\mathbb C$ and $\HQ$, the
  primitive data for contiguous datastrutures which contain 
  elements of type $\mathbb F$ should be arranged into a data 
  layout which allows for a minimum number of MOPs to be performed
  from within the microkernel, as long as map between the standard
  and new data layout is space preserving.
\end{itemize}
To demonstrate what is meant by a space preserving map in this
context, consider an complex element, $z = a + bi \in \mathbb C$,
which is represented by two primitive real numbers $a,b \in
\mathbb R$ which are contiguous in memory, denoted $[a;b]$. For a
datastructure which contains two contiguous elements
$z_1,z_2\in\mathbb C$, the data layouts $[a_1;b_1;a_2;b_2]$ and
$[a_1;a_2;b_1;b_2]$ occupy the same space in memory. Thus a map
between these two data layouts would be considered space
preserving.  While the first two aspects of data packing are well
explored in the literature, the latter has not to the best of
authors' knowledge. As will be demonstrated in
\cref{sec:matmult_quaternion}, optimizing the primitive data
layout of contiguous quaternion datastructures will prove
important in the development of an optimized quaternion GEMM.

\begin{algorithm2e}[t]
  \caption{Abstract Template for the \texttt{PACK1} Kernel}
  \label{alg:pack1_abstract}
  \SetKwInOut{kinput}{Input}
  \SetKwInOut{koutput}{Output}
  \BlankLine
  \kinput{Identified sub-matrix $A_r \in \MG{M_c}{K_c}{F}$ (non-contiguous)}
  \BlankLine
  \koutput{Packed sub-matrix $\tilde A_p \in \MG{K_cM_r}{M_c/M_r}{F}$ (contiguous)}
  \BlankLine
  \BlankLine

  \For{$\mu=1:M_c:M_r$}{
  \For{$\kappa=1:K_c$}{
    \nl $i \leftarrow M_r(\kappa-1) + 1$\\
    \nl $\tilde A_p([i,i+M_r],\mu/M_r) \leftarrow \texttt{PACKOP1}(A_r([\mu,\mu+M_r],\kappa))$\\
  }
  }
\end{algorithm2e}

\begin{algorithm2e}[b]
  \caption{Abstract Template for the \texttt{PACK2} Kernel}
  \label{alg:pack2_abstract}
  \SetKwInOut{kinput}{Input}
  \SetKwInOut{koutput}{Output}
  \BlankLine
  \kinput{Identified sub-matrix $B_r \in \MG{K_c}{N_c}{F}$ (non-contiguous)}
  \BlankLine
  \koutput{Packed sub-matrix $\tilde B_p \in \MG{K_cN_r}{N_c/N_r}{F}$ (contiguous)}
  \BlankLine
  \BlankLine

  \For{$\nu=1:N_c:N_r$}{
  \For{$\kappa=1:K_c$}{
    \nl $i \leftarrow N_r(\kappa-1) + 1$\\
    \nl $\tilde B_p([i,i+N_r],\nu/N_r) \leftarrow \texttt{PACKOP2}(B_r(\kappa,[\nu,\nu+N_r]))$\\
  }
  }
\end{algorithm2e}
The fact that the rank-1 updates required by
\cref{eq:micro_kernel,alg:abstract_kern} involve both row and
column vectors, a single packing strategy would not be sufficient
to achieve optimal data layout for both $\tilde A_p$ and $\tilde
B_p$. Thus, the packing kernels \texttt{PACK1} and \texttt{PACK2}
must be designed separately to optimize the layouts of $\tilde
A_p$ and $\tilde B_p$, respectively. An abstract templates for
these packing kernels are given in
\cref{alg:pack1_abstract,alg:pack2_abstract}, respectively. We
refer the reader to the work of Van Zee, \emph{et al} \cite{BLIS1}
for an intuitive graphical illustration of the optimal packing
procedure.  To account for the rearrangement of primitive data in
the packing procedure, we have introduced two additional
operations, \texttt{PACKOP1} and \texttt{PACKOP2}, to perform this
operation for the kernels \texttt{PACK1} and \texttt{PACK2},
respectively.  Note that typical implementations for real and
complex GEMM would yield both \texttt{PACKOP1} and
\texttt{PACKOP2} as either the identity or linear scaling
operation. 

Due to the large access latency difference between RAM and the
other levels of the memory hierarchy, operations performed within
\texttt{PACKOP1} and \texttt{PACKOP2} have little to no impact on
the performance of the GEMM implementation.  This is due to the
fact that that these operations are to be done in the registers,
and are thus amortized over the time it takes to access the data
from the RAM. For example, the construction of the packed
sub-matrix $\tilde A_p$ in \cref{alg_ln:a_pack} of
\cref{alg:opt_gemm} requires the scaling of the sub-matrix
$\subrowcol{A}{\mu}{\kappa} \rightarrow \alpha
\subrowcol{A}{\mu}{\kappa}$. As there is a two orders of magnitude
latency ratio between RAM access (O(100s) of clock cycles) and the
FLOP required to scale an element of the matrix (O(4-5) clock
cycles), the cost of the scaling operation may be thought of as
negligible. The same logic holds true for data rearrangement
operations, such as register transpose, which will be explored in
the following section.

\section{Quaternion Matrix Multiplication: HGEMM}
\label{sec:matmult_quaternion}

In this section, we develop the details of a high--performance
implementation of quaternion GEMM for the AVX microarchitechture.
The primary focus of this section is the development of
AVX-optimized versions of the kernels described in the previous
section for use with quaternion arithmetic and data structures. In
practice, there are two primary features of the AVX
microarchitechture that one must consider in the development of
optimized GEMM kernels:
\begin{enumerate}
  \item processors with support for AVX instructions have (at least) 16 256-bit floating point (\texttt{YMM}) registers, and
  \item AVX dictates support for SIMD (but not FMA) arithmetic instructions on these \texttt{YMM} registers. 
\end{enumerate}
For the purposes of this work, we will restrict the discussion of
kernel development to double precision floating point storage,
i.e. each floating point primitive will occupy 64-bits. As such,
each \texttt{YMM} register on AVX can hold and perform arithmetic
operations on up to 4 double precision floats, simultaneously.  In
analogy to the DGEMM and ZGEMM naming conventions of real and
complex GEMM operations, we will refer to the double precision
quaternion GEMM as HGEMM.  As an extension of the standard
construction of complex datatypes as two contiguous floats, the
following developments will describe quaternion datatypes as four
contiguous floats, $[q^0; q^1; q^2; q^3]$ using the notation of
\cref{eq:quaternion_def}.  As such, each AVX \texttt{YMM} register
can hold one double precision quaternion (or equivalent) at any
point in time.

\subsection{Batch SIMD Quaternion Multiplication}
\label{sec:quaternion_simd}

Critical to the development of an AVX-optimized quaternion
microkernel is an efficient strategy for quaternion product using
SIMD arithmetic operations. The the product of quaternions given
by the Hamilton product in \cref{eq:quaternion_prod} requires a
minimum of 16 FLOPs to complete. As each \texttt{YMM} register in
AVX is capable of storing and manipulating 4 floats at once, one
could in principle perform some of these FLOPs concurrently if the
task is simply to perform a single quaternion product.
However, if the task is to perform many quaternion products in a
structured manner, as is the case for the rank-1 updates required
by \cref{eq:micro_kernel}, implementations which optimize for a
single quaternion product will yield sub-optimal throughput. To
leverage the full power of SIMD instructions in this case, one
needs to develop a strategy which aims to perform multiple
quaternion products simultaneously at the highest throughput
possible. As each \texttt{YMM} register is able to manipulate 4
floats, the simplest manner to reach optimal throughput is to
perform 4 quaternion products simultaneously.

Consider the batch quaternion product which takes two sets of four
quaternions, $\{p_i\}_{i=1}^4$ and $\{q_i\}_{i=1}^4$, and returns
a set of four quaternion products, $\{(pq)_i\}_{i=1}^4$.  For
simplicity in the following, we will augment the product operation
to perform an update of the result as opposed to an assignment, 
\begin{equation}
\begin{bmatrix} (pq)_1 \\ (pq)_2 \\ (pq)_3 \\ (pq)_4 \end{bmatrix} = 
\begin{bmatrix} (pq)_1 \\ (pq)_2 \\ (pq)_3 \\ (pq)_4 \end{bmatrix} +
\begin{bmatrix} p_1 \\ p_2 \\ p_3 \\ p_4 \end{bmatrix} \circ
\begin{bmatrix} q_1 \\ q_2 \\ q_3 \\ q_4 \end{bmatrix} 
\label{eq:simd_batch_mul}
\end{equation}
where $\circ$ is the Hadamard (entry-wise) product such that
$(pq)_1 = (pq)_1 + p_1q_1$ and so on.  Each quaternion product
(\cref{eq:quaternion_prod}) may be separated into 4 sets of 4
FLOPs which update each component of the result, respectively. In
the following, we examine the update of the scalar component of
the product, $(pq)_i^0$, as a representative example.  Note that
extension and generalization to vector components of $(pq)_i$ is
straightforward.  The scalar components of each updated quaternion
product may be obtained simultaneously by
\begin{equation}
\begin{bmatrix} (pq)^0_1 \\ (pq)^0_2 \\ (pq)^0_3 \\ (pq)^0_4 \end{bmatrix} = 
\begin{bmatrix} (pq)^0_1 \\ (pq)^0_2 \\ (pq)^0_3 \\ (pq)^0_4 \end{bmatrix} +
\begin{bmatrix} p^0_1 \\ p^0_2 \\ p^0_3 \\ p^0_4 \end{bmatrix} \circ
\begin{bmatrix} q^0_1 \\ q^0_2 \\ q^0_3 \\ q^0_4 \end{bmatrix} -
\begin{bmatrix} p^1_1 \\ p^1_2 \\ p^1_3 \\ p^1_4 \end{bmatrix} \circ
\begin{bmatrix} q^1_1 \\ q^1_2 \\ q^1_3 \\ q^1_4 \end{bmatrix} -
\begin{bmatrix} p^2_1 \\ p^2_2 \\ p^2_3 \\ p^2_4 \end{bmatrix} \circ
\begin{bmatrix} q^2_1 \\ q^2_2 \\ q^2_3 \\ q^2_4 \end{bmatrix} -
\begin{bmatrix} p^3_1 \\ p^3_2 \\ p^3_3 \\ p^3_4 \end{bmatrix} \circ
\begin{bmatrix} q^3_1 \\ q^3_2 \\ q^3_3 \\ q^3_4 \end{bmatrix}. 
\label{eq:simd_batch_mul_scalar}
\end{equation}
In the SIMD paradigm, each of these vectors may be represented by
a single \texttt{YMM} register. As such, each of these Hadamard
products may be performed by the \texttt{VMULPD} vector
instruction and each vector addition (subtraction) by the
\texttt{VADDPD} (\texttt{VSUBPD}) vector instruction. In this
form, the entire batch quaternion multiplication may be completed
using 32 vector instructions.

\begin{figure}
  \centering
  \includegraphics[width=0.5\textwidth]{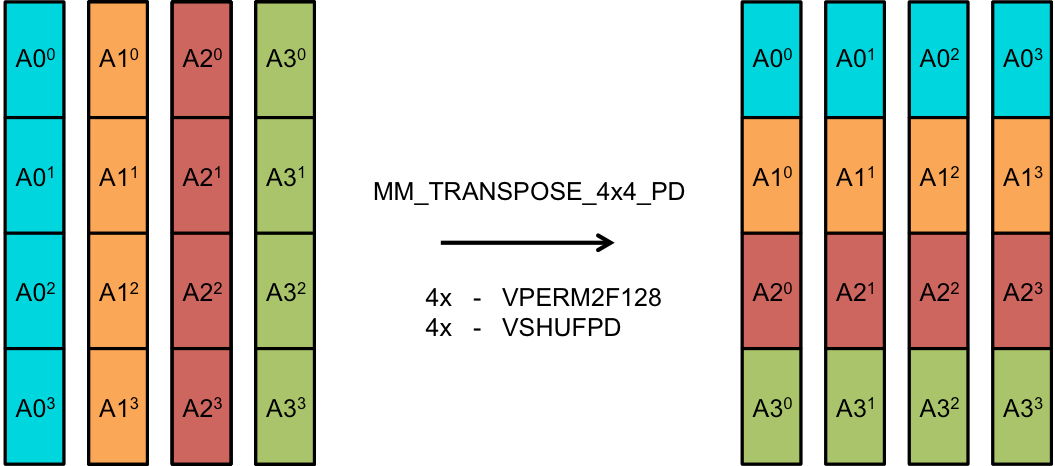}
  \caption{Graphical illustration of a 4x4 register transpose of 4 \texttt{YMM} registers containing general
  contiguous data structures. In the general case, this operation may be completed using 4x \texttt{VPERM2F128}
  and 4x \texttt{VSHUFPD} vector instructions and 4 additional \texttt{YMM} registers which may be used as scratch space.}
  \label{fig:register_trans}
\end{figure}
The structure of \cref{eq:simd_batch_mul_scalar} requires that
each of the sets $\{p_i\}$, $\{q_i\}$ and $\{(pq)_i\}$ occupy 4
\texttt{YMM} registers, with each register containing a particular
quaternion component of each element in the set, respectively. In
other words, one \texttt{YMM} register contains all of the scalar
components for each element of $\{p_i\}$, one for the scalar
components of $\{q_i\}$, and so on for the vector components of
these sets and for the components of $\{(pq)_i\}$. For clarity in
the following, we will denote the \texttt{YMM} register containing
the scalar components of $\{p_i\}$, $\{q_i\}$ and $\{(pq)_i\}$ as
$\texttt{P}^0$, $\texttt{Q}^0$ and $\texttt{PQ}^0$, respectively,
and so on for the vector parts of these sets with indices $1$, $2$
and $3$. Using this notation, we will define the SIMD
implementation of \cref{eq:simd_batch_mul} as 
\begin{equation}
(\texttt{PQ}^0,\texttt{PQ}^1,\texttt{PQ}^2,\texttt{PQ}^3) \leftarrow 
\texttt{HMUL}(\{\texttt{PQ}^i\},\{\texttt{P}^i\},\{\texttt{Q}^i\}). \label{eq:simd_batch_mul_impl}
\end{equation}
For quaternion data structures which store a single quaternion
contiguously, such as the one considered in this work, the vector
load instruction (\texttt{VMOVAPD}) would populate each register
with the 4 components of a single quaternion. As such, one would
need to rearrange the quaternion data once it is read into
registers in order to utilize \cref{eq:simd_batch_mul_impl}. In
general, this rearrangement may be achieved by a 4x4 register
transpose on each of the quaternion sets. This register transpose
will be denoted \texttt{MM\_4x4\_TRANSPOSE\_PD} in the following
and  is illustrated graphically in \cref{fig:register_trans}. For
clarity, we endow \texttt{MM\_4x4\_TRANSPOSE\_PD} with the
function signature
\begin{equation}
(\texttt{P}^0,\texttt{P}^1,\texttt{P}^2,\texttt{P}^3) \leftarrow 
\texttt{MM\_4x4\_TRANSPOSE\_PD}(\texttt{P}_1,\texttt{P}_2,\texttt{P}_3,\texttt{P}_4), \label{eq:gen_register_trans_impl}
\end{equation}
where $\texttt{P}_1$ is a \texttt{YMM} register containing the
components of $p_1$, $\texttt{P}_2$ the components of $p_2$, and
so on. Remark that the result of \texttt{MM\_4x4\_TRANSPOSE\_PD}
is \emph{not} invariant to the permutation of its parameters.
Further, we note that \texttt{MM\_4x4\_TRANSPOSE\_PD} is an
involution.  In general, register transpose is a relatively
expensive operation due to the high aggregate latency of the
vector instructions (\texttt{VPERM2F128} and \texttt{VSHUFPD})
involved in its implementation. However, it will be shown in the
following subsection that the special structure of the rank-1
update will simplify and cheapen the general register transpose
through the use of optimal packing layouts in the GEMM operation.

\subsection{The Quaternion Microkernel and Amortization of Register Transpose}
\label{sec:quaternion_kern}

Given that AVX only supports 16 \texttt{YMM} registers, the
largest register block (\cref{eq:micro_kernel}) which allows for
$C_r$, $\singcol{A_r}{\kappa}$ and $\singrow{B_r}{\kappa}$ to 
all reside in registers simultaneously is given by
$N_r=M_r=2$. As such, the quaternion microkernel must perform a
sum over 2x2 rank-1 updates to update a register block of $C$. A
single 2x2 rank-1 update requires 4 product evaluations given by
\begin{equation}
\begin{bmatrix} C_{r,11} \\ C_{r,12} \\ C_{r,21} \\  C_{r,22} \end{bmatrix} =  
\begin{bmatrix} C_{r,11} \\ C_{r,12} \\ C_{r,21} \\  C_{r,22} \end{bmatrix} +  
\begin{bmatrix} \vc a_{r,1}  \\ \vc a_{r,1}  \\ \vc a_{r,2}  \\  \vc a_{r,2} \end{bmatrix} \circ 
\begin{bmatrix} \vc b_{r,1}  \\ \vc b_{r,2}  \\ \vc b_{r,1}  \\  \vc b_{r,2} \end{bmatrix},
\label{eq:quaternion_kern_rankupdate}
\end{equation}
where we have dropped the $(\kappa)$ super-- and subscripts for
brevity.  Per the discussion of the previous subsection, these
product evaluations may be performed simultaneously using SIMD
vector instructions given that the register data arrangement
adheres to the structure \cref{eq:simd_batch_mul_impl} via
\cref{eq:gen_register_trans_impl}.  On top of the 32 vector
instructions requires to perform the product accumulations, the
general scheme for register transpose depicted in
\cref{fig:register_trans} requires an additional 16 register
operations: 8 for transposing the components of $\vc a_r$ and
$\vc b_r$, respectively.  The operation overhead is further
compounded by the fact that the microkernel performs many ($K_c$)
rank-1 updates successively, thus this scheme costs $16K_c$
additional operations over the execution of the microkernel.
However, such a general approach for register transpose would only
be required for 4 \emph{unique} quaternions, whereas the 4
(unique) products required for the evaluation of
\cref{eq:quaternion_kern_rankupdate} only involve 2 sets of 2
unique quaternions. As such, simplifications to the general
register transpose scheme of \cref{fig:register_trans} may be made
in this case.

\begin{figure}
  \centering
  \begin{subfigure}[b]{0.46\textwidth}
    \centering
    \includegraphics[width=\textwidth]{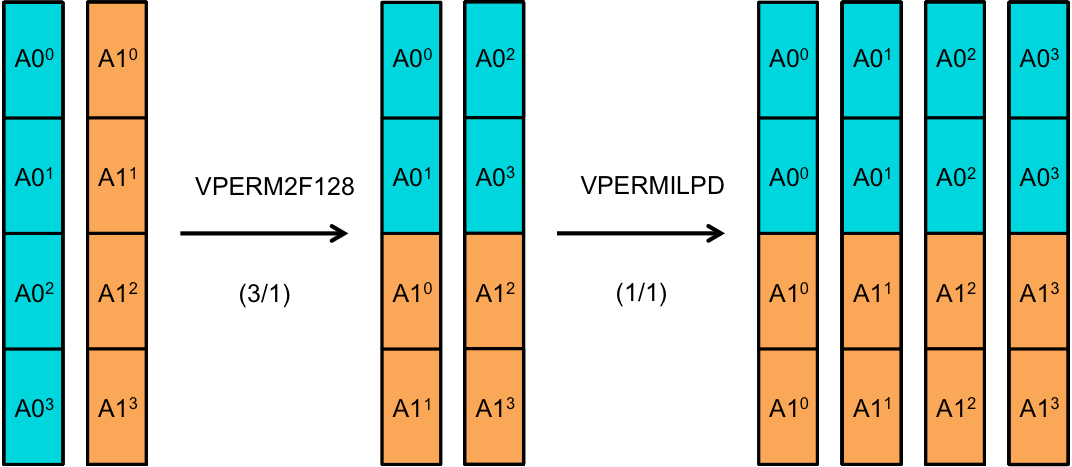}
    \caption{}
    \label{fig:atrans}
  \end{subfigure}~$\qquad$
  \begin{subfigure}[b]{0.46\textwidth}
    \centering
    \includegraphics[width=\textwidth]{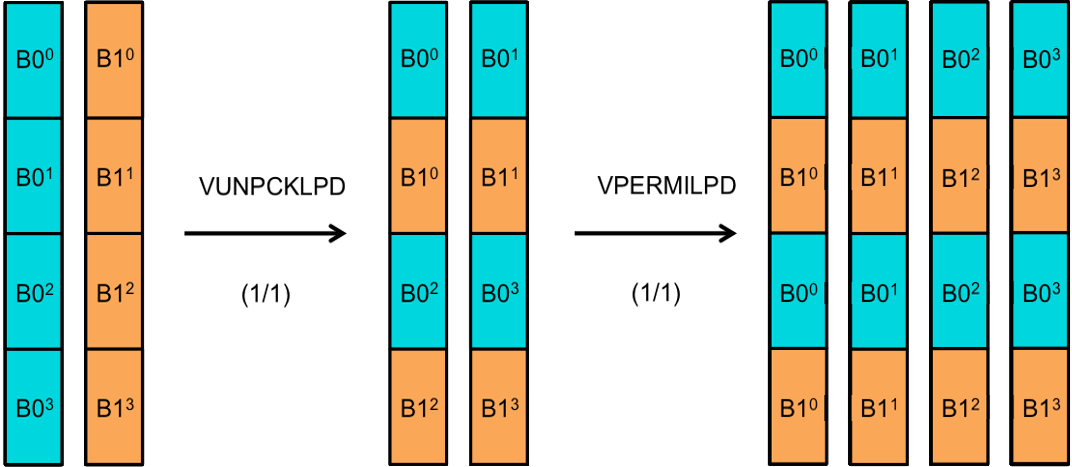}
    \caption{}
    \label{fig:btrans}
  \end{subfigure}
  \caption{Alternative register transpose schemes to efficiently handle redundancies in the general
  4x4 scheme depicted in \cref{fig:register_trans}. \Cref{fig:atrans} handles the transpose case for the
  elements of $\singcol{A_r}{\kappa}$, and \cref{fig:btrans} for the elements of $\singrow{B_r}{\kappa}$.
  Both of these schemes decompose the general register transpose operation into two operations; the first of 
  which is space preserving. The labels beneath the arrows indicate the latency / reciprocal throughput
  for the instruction on the Intel(R) Sandy-Bridge microarchitechture.
  }
  \label{fig:alt_trans}
\end{figure}
There are two special cases for register transpose which we must
consider for \cref{eq:quaternion_kern_rankupdate}, namely those
which represent the data ordering of the elements of $\vc{A_r}$
and $\vc{B_r}$, respectively:
\begin{align}
  (\texttt{A}^0,\texttt{A}^1,\texttt{A}^2,\texttt{A}^3) &\leftarrow 
  \texttt{MM\_4x4\_TRANSPOSE\_PD}(\texttt{A}_1,\texttt{A}_1,\texttt{A}_2,\texttt{A}_2), \\ 
  (\texttt{B}^0,\texttt{B}^1,\texttt{B}^2,\texttt{B}^3) &\leftarrow 
  \texttt{MM\_4x4\_TRANSPOSE\_PD}(\texttt{B}_1,\texttt{B}_2,\texttt{B}_1,\texttt{B}_2). 
\end{align}
Here, $\texttt{A}_1$ and $\texttt{A}_2$ hold the components of
$\vc a_{r,1}$ and $\vc a_{r,2}$, respectively, and similarly for
$\texttt{B}_1$ and $\texttt{B}_2$ for the elements of $\vc b_r$.
The \texttt{YMM} registers $\{\texttt{A}^i\}$ and
$\{\texttt{B}^i\}$ represent the components of $\vc a_r$ and
$\vc b_r$ in the order which they were passed, i.e.
$\texttt{A}^0$ has the layout
$[\vc{a}_{r,1}^0;\vc{a}_{r,1}^0;\vc{a}_{r,2}^0;\vc{a}_{r,2}^0 ]$ while
$\texttt{B}^0$ has the layout
$[\vc{b}_{r,1}^0;\vc{b}_{r,2}^0;\vc{b}_{r,1}^0;\vc{b}_{r,2}^0 ]$ and
so on.  The presence of redundancies in the register transpose
allows for factorization of \texttt{MM\_4x4\_TRANSPOSE\_PD} into
the convolution of two simpler operations:
\begin{align}
  (\texttt{A}^0,\texttt{A}^1,\texttt{A}^2,\texttt{A}^3) &\leftarrow 
  \texttt{ATRANS2}(\texttt{ATRANS1}(\texttt{A}_1,\texttt{A}_2)), \\ 
  (\texttt{B}^0,\texttt{B}^1,\texttt{B}^2,\texttt{B}^3) &\leftarrow 
  \texttt{BTRANS2}(\texttt{BTRANS1}(\texttt{B}_1,\texttt{B}_2)). 
\end{align}
An illustration of this state of affairs is given in
\cref{fig:alt_trans} with \cref{fig:atrans,fig:btrans} depicting
the transpose of elements of $\vc a_r$ and $\vc b_r$,
respectively. The first step of \cref{fig:atrans} demonstrates the
effect of \texttt{ATRANS1} and the second the effect of
\texttt{ATRANS2}, and similarly for \cref{fig:btrans}. The most
important aspect of the alternative transpose schemes depicted in
\cref{fig:alt_trans} is that both \texttt{ATRANS1} and
\texttt{BTRANS1} are space preserving. As such, they may be
factored into the packing scheme as discussed in
\cref{sec:packing}, leading to an amortization of register
operations over data movement from RAM. Further, in the case of
\texttt{ATRANS1}, not only does this procedure reduce the number
of instructions which must be issued from inside the microkernel
loop, it does so in a way that the most \emph{expensive} (highest
latency) register operations required for the transpose are
amortized in the packing procedure. In the context of
\cref{alg:pack1_abstract,alg:pack2_abstract}, this factorization
may be accounted for by setting $\texttt{PACKOP1}=\alpha *
\texttt{ATRANS1}$ and $\texttt{PACKOP2} = \texttt{BTRANS1}$.
Utilizing this packing strategy, the operation overhead for
performing register transpose from within the microkernel is
reduced by a factor of 3/4 ($16K_c$ to $4K_c$).
\Cref{alg:avx_microkernel} outlines the general structure for the AVX optimized HGEMM microkernel. The following
section demonstrates its performance.

\begin{algorithm2e*}[t]
  \caption{AVX Optimized HGEMM Microkernel ($N_r=M_r=2$)}
  \label{alg:avx_microkernel}
  \SetKwInOut{kinput}{Input}
  \SetKwInOut{koutput}{Output}
  \kinput{ Columns of packed matrices $\tilde a_p \in \VG{2K_c}{H}$, 
          $\tilde b_p \in \VG{2K_c}{H}$,\\
          Register block $C_r \in \MGsq{2}{H}$ of $C$.  }
  \koutput{Updated $C_r$}

  \BlankLine
  \nl Stream $C_{r,11},C_{r,12},C_{r,21},C_{r,22}$ into registers 
       $\texttt R_{11},\texttt R_{12},\texttt R_{21},\texttt R_{22}$ from RAM\\
  \nl ($\texttt R^0, \texttt R^1, \texttt R^2, \texttt R^3) \leftarrow $ \texttt{MM\_TRANSPOSE\_4x4\_PD}
       $(\texttt R_{11},\texttt R_{12},\texttt R_{21},\texttt R_{22})$\\

  \BlankLine
  \For{$k = 1:k_c$}{

    \BlankLine
    \nl Load $\texttt A_1 \leftarrow \vc{a}_{p,2k}$, $\texttt A_2 \leftarrow \vc{a}_{p,2k+1}$\\
    \nl Load $\texttt B_1 \leftarrow \vc{b}_{p,2k}$, $\texttt B_2 \leftarrow \vc{b}_{p,2k+1}$\\
    \nl ($\texttt A^0, \texttt A^1, \texttt A^2, \texttt A^3) \leftarrow $ \texttt{ATRANS2}
       $(\texttt A_1,\texttt A_2)$\\
    \nl ($\texttt B^0, \texttt B^1, \texttt B^2, \texttt B^3) \leftarrow $ \texttt{BTRANS2}
       $(\texttt B_1,\texttt B_2)$\\
    \nl $(\texttt R^0, \texttt R^1, \texttt R^2, \texttt R^3) \leftarrow$
      $\texttt{HMUL}(\{\texttt R^i\}, \{\texttt A^i\}, \{\texttt B^i\})$ \\
    \BlankLine

  }
  \BlankLine
  \nl ($\texttt R_{11},\texttt R_{12},\texttt R_{21},\texttt R_{22}) \leftarrow $ \texttt{MM\_TRANSPOSE\_4x4\_PD}
       $(\texttt R^0, \texttt R^1, \texttt R^2, \texttt R^3)$\\
  \nl Store $\texttt R_{11},\texttt R_{12},\texttt R_{21},\texttt R_{22}$ in $C_{r,11},C_{r,12},C_{r,21},C_{r,22}$

\end{algorithm2e*}

\section{Implementation and Performance Results}
\label{sec:results}

HGEMM, as described in the previous section, has been implemented
in the quaternion BLAS (HBLAS) component of the HAXX library. HAXX
({\bf H}amilton's Quaternion {\bf A}lgebra for C{\bf XX})
\cite{HAXX} is a modern C++ software infrastructure developed to
enable efficient scalar and linear algebra operations using
quaternion and mixed--type (quaternion--complex, quaternion--real)
arithmetic.  As of this work, HAXX provides reference and
optimized serial implementations for a representative subset of
BLAS-1,2,3 functionality. For the optimized implementation of
HGEMM in HAXX, the arithmetic microkernel has been implemented
using C++ vector--intrinsics rather than the assembly
implementations which have become ubiquitous in high--performance
implementation of DGEMM and ZGEMM. This has been done primarily
for the fact that vector intrinsics offer a reasonable balance
between transparency in the code-base and potential performance
from low-level access to assembly instructions, even if this
transparency comes at a slight performance degradation.  In this
section, we provide performance results for the reference and
optimized HGEMM implementations in HAXX for the AVX
microarchitechture. All timing results were obtained using an
Intel(R) Xeon(R) CPU E5-2660 (Sandy Bridge) @ 2.20 GHz (max 3.0
GHz). The E5-2660 processor admits cache sizes of 32 kB, 256 kB
and 20 MB for the L1, L2, and L3 caches respectively. The L3 cache
is shared among all cores on the CPU. Theoretical (serial)
peak performance double precision arithmetic on this CPU is 24
GFLOP/s. HAXX and all benchmark executables were compiled using
the Intel(R) C++ compiler with architecture specific
optimizations (`\texttt{-xHost}') and interprocedural optimization
enabled.  To obtain the caching parameters, the open--source
autotuning software OpenTuner \cite{OpenTuner} was employed. On
this architecture, the optimal caching parameters were found to be
$M_c=N_c=64$ and $K_c=1024$.

\begin{figure}[t]
  \centering
  \begin{subfigure}[b]{0.48\textwidth}
    \centering
    \includegraphics[width=\textwidth]{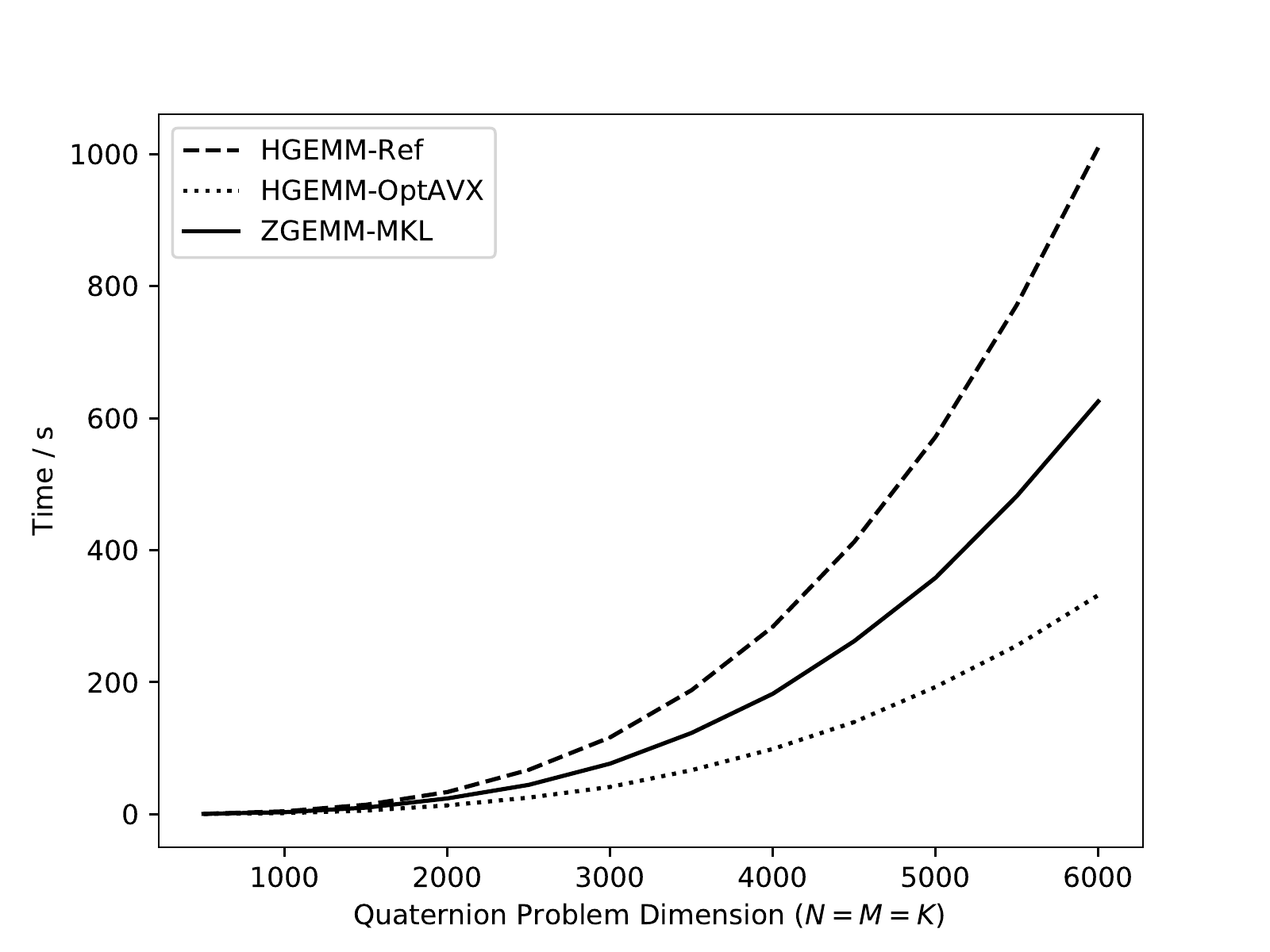}
    \caption{}
    \label{fig:hgemm_time}
  \end{subfigure}
  \begin{subfigure}[b]{0.48\textwidth}
    \centering
    \includegraphics[width=\textwidth]{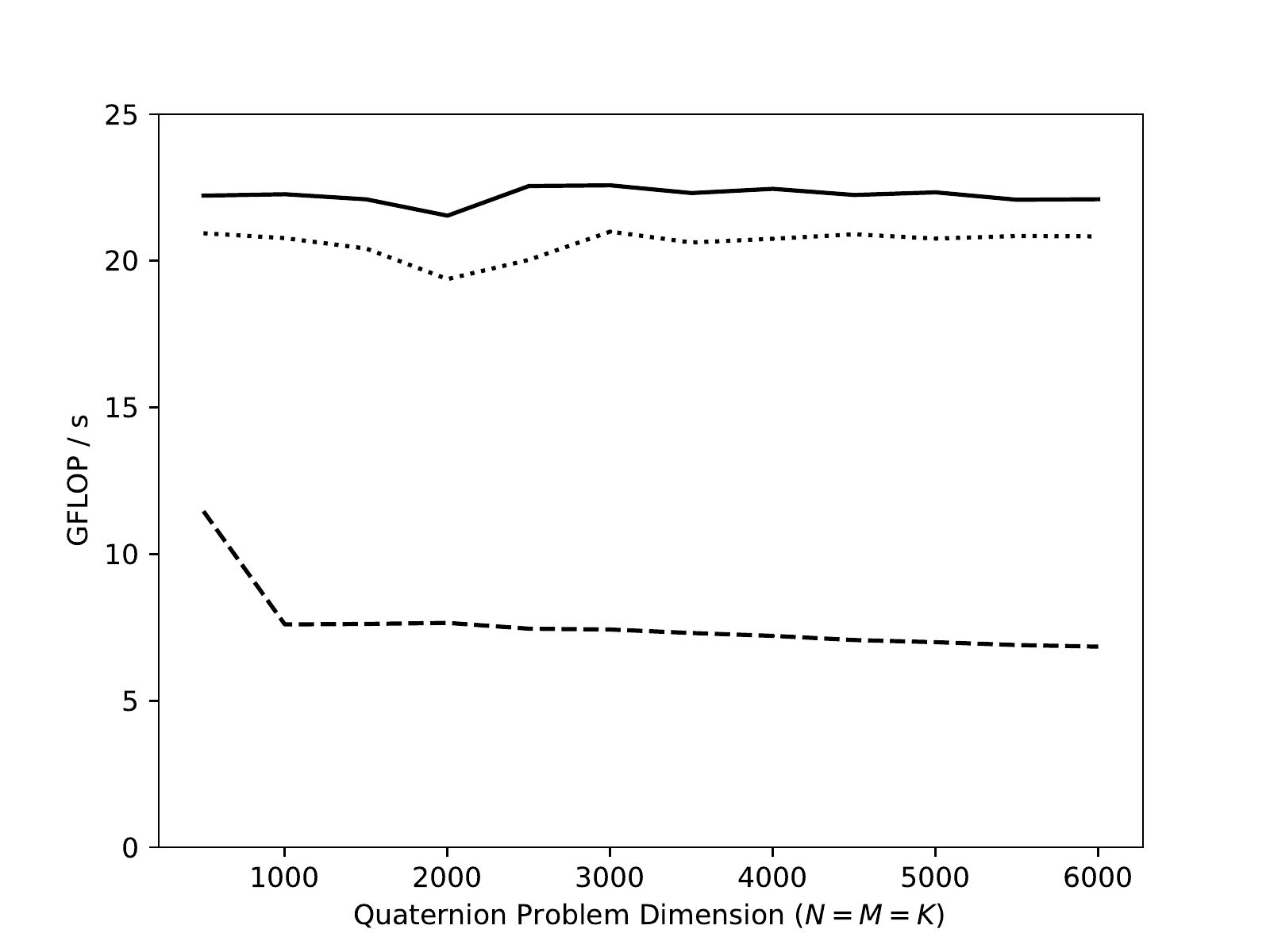}
    \caption{}
    \label{fig:hgemm_flop}
  \end{subfigure}
  \caption{Computational timing and scaling comparisons for 
  reference HGEMM (HGEMM-Ref), AVX optimized HGEMM (HGEMM-OptAVX) 
  and the serial ZGEMM of Intel MKL (ZGEMM-MKL).
  \Cref{fig:hgemm_time} shows the raw timing 
  comparisons and
  \cref{fig:hgemm_flop} shows the FLOP rate comparisons
  between the GEMM implementations. 
  }
  \label{fig:hgemm_zgemm_comp}
\end{figure}

\Cref{fig:hgemm_zgemm_comp} illustrates performance comparisons
for three GEMM implementations: the reference (HGEMM-Ref) and
AVX-optimized (HGEMM-OptAVX) HGEMM implementations provided in the
HAXX library, and the (serial) ZGEMM implementation provided by
the Intel(R) Math Kernel Library (MKL) (Version 18.0 Update 1). 
All timings presented are representative of
\cref{eq:general_gemm} for square matrices with $\alpha=1$ and
$\beta=0$. Timings for the HGEMM implementations are for the
matrix product operation on $\MGsq{N}{H}$ while those for ZGEMM are
for the analogous product operation on $\MGsq{2N}{C}$ (see
\cref{eq:hc_mat_mult}).  The comparison between complex and
quaternion operations are presented in this manner to demonstrate
the efficacy of the quaternion operation over the complex
operation for \emph{the same arithmetic operation}, i.e. the
results of these operations represent the same mathematical object
(up to an isomorphism). There two primary results which are to be
taken from these numerical experiments:
\begin{enumerate}
  \item The timing comparisons depicted in \cref{fig:hgemm_time}
  illustrate that quaternion arithmetic alone is not sufficient to
  obtain performance leverage over tuned complex matrix
  multiplication. The reference HGEMM implementation is
  significantly less performant than the ZGEMM implementation found
  in MKL, while the AVX optimized HGEMM implementation outperforms
  the ZGEMM operation by roughly a factor of 2 (as would be expected
  from \cref{tbl:hc_mat_flops}).  Further, as was described in
  \cref{sec:matmult}, the reference HGEMM implementation performs
  significantly under the theoretical peak performance
  (\textasciitilde7 GFLOP/s vs 24 GFLOP/s) due to the algorithm
  being memory bound.

  \item Despite a slight difference in the FLOP rate in the 
  GEMM implementations depicted in \cref{fig:hgemm_flop}
  (\textasciitilde22 GFLOP/s for ZGEMM-MKL and 
  \textasciitilde21 GFLOP/s for HGEMM-OptAVX),
  the optimized HGEMM implementation consistently outperforms
  the optimized ZGEMM implementation even for large matrices
  ($N > 3000$).
\end{enumerate}

%\todo{Does this need to be said? It may be getting into the weeds}
%
%There are a number of possible sources for the FLOP rate
%differences between the ZGEMM-MKL and HGEMM-OptAVX implementation.
%The most likely scenario for this performance difference is in the
%method that the microkernel has been implementation between the
%two GEMM implementations. It is common practice \todo{cite?} for
%one to unroll the inner-loop contained in the microkernel to
%minimize the the number of times one has to increment the loop
%counter and thus reduce the number of operations which must be
%performed from within the microkernel. Such a feature is likely
%present in the MKL implementation of ZGEMM \todo{we know this is
%true, but how to explain it with the proprietary nature of MKL?}.
%However, in the vector-intrinsics implementation of the HGEMM
%microkernel in HGEMM-OptAVX, loop unrolling has not been used as
%its utilization typically yields sub-optimal performance due to
%compiler optimization heuristics.

\section{Conclusions}
\label{sec:conclusions}

In this work, we have demonstrated the efficacy and potential of
high--performance quaternion linear algebra to leverage performance
increases over complex linear algebra for special class of
matrices through the efficient implementation of the quaternion
matrix product. The software development proposed in this work
extends the existing theory of high-performance serial matrix
multiplication for use with explicitly quaternion arithmetic,
as outlined in \cref{sec:matmult,sec:matmult_quaternion}. A series
of numerical experiments given in \cref{sec:results} have
illustrated performance comparisons between reference quaternion,
optimized quaternion, and vendor tuned complex GEMM
implementations. It was shown that exploitation of quaternion
arithmetic alone is not sufficient to outperform high-performance
implementations of complex GEMM and that analogous implementations
of high-performance quaternion GEMM are necessary to
leverage such improvements. Further, it was shown that even
in the presence of slight difference the FLOP rate comparisons,
the optimized implementation of quaternion GEMM outperforms 
the optimized implementation of complex GEMM for the analogous 
arithmetic operation. 
We note for completeness that while Intel(R) Sandy Bridge and the AVX
instruction set are not contemporary in and of themselves, they representative
of more contemporary architectures such as the Intel(R) Haswell and AMD(R)
Excavator architectures which support the AVX2 instruction set. In the context
of the GEMM operation, the primary feature introduced in these architectures
is FMA arithmetic instructions. With the exception of architectures which support the
AVX-512 instruction set (such as Intel(R) Skylake-X and Intel(R) Knight's
Landing), the SIMD vector units on architectures which support either the AVX or AVX2
instruction sets are 256 bits in length.  Thus with the exception of the 
arithmetic kernel (\cref{eq:simd_batch_mul_impl}) and the optimal values of the
caching parameters, the remainder of the findings in this work would would be
invariant between AVX and AVX2.
In summary, the optimized implementation
of quaternion GEMM provided by the HAXX library was shown
to outperform its MKL optimized complex analogue by roughly a 
factor of 2 (as would be expected from the discussion in 
\cref{sec:quaternion_la}). As the architecture on which
the numerical experiments were performed is a representative
example of modern HPC architectures in general, the results
presented in this work would translate to other architectures
given that one provides optimized versions of the GEMM kernels for
the architecture in question.

As the matrix product is the fundamental building
block for the development of important operations such as
eigendecomposition and matrix factorization, its efficient
implementation is a necessary condition for high-performance
linear algebra software.  In order for quaternion linear algebra
to be a viable alternative to complex linear algebra in problems
which it may be applied, optimized implementations of quaternion
operations which outperform their complex counterparts must be
developed.  Although the power of the \emph{theory} of quaternion
algebra in the context of scientific theory and computation has
been known for some time, prior to this work, no performant
implementation of quaternion linear algebra has been available.
It is our hope that the software developments presented in this
work will aid and spark interest in the future development of
high-performance quaternion linear algebra such that the full
power of quaternion arithmetic may be leveraged in computationally
intensive fields such as scientific computing and image processing.

\begin{acks}
In the development of HAXX, DWY was supported by a fellowship from The
Molecular Sciences Software Institute under National Science Foundation grant
ACI-1547580. The development of HAXX has also been supported through the
development of the open source Chronus Quantum supported by the National
Science Foundation (OAC-1663636 to XL). This work has been further supported in
part by the U.S. Department of Energy, Office of Science, Basic Energy
Sciences, under Award LAB 17-1775, as part of the Computational Chemical
Sciences Program.

The authors would like to thank Edward Valeev and Benjamin Pritchard for
insightful discussions regarding microarchitechture optimizations and
high-performance matrix multiplication and Wissam Sid Lakhdar for aid in the
tuning of HAXX. Further, the authors would like to thank Benjamin Pritchard,
Wissam Sid Lakhdar, Joseph Kasper and the anonymous reviewers
for reviewing the content of the manuscript and providing meaningful insight.
\end{acks}

\bibliographystyle{ACM-Reference-Format}
\bibliography{Journal_Short_Name,quaternions_in_qc,relativistic_qc,qc_software,quaternion_refs,quaternions_image_signal_processing,linalg_ref,misc_qc_refs,Li_Group_References,misc_refs}

\end{document}